\documentclass[11pt]{article}

\usepackage{amsfonts}
\usepackage{amsmath}
\usepackage{amssymb}
\usepackage{amsthm}
\usepackage[mathscr]{eucal}
\usepackage{bbold}
\usepackage{bm}
\usepackage{braket}
\usepackage{color}
\usepackage{comment}
\usepackage{dsfont}
\usepackage{framed}
\usepackage{jheppub}
\usepackage{mathtools}
\usepackage{multirow}
\usepackage{physics}
\usepackage[normalem]{ulem}
\usepackage{tensor}
\usepackage{thmtools}
\usepackage{thm-restate}
\usepackage{enumitem}
\usepackage{pifont}
\usepackage{ulem}
\usepackage{tikz}
\usetikzlibrary{math}
\usepackage{graphicx}
\usepackage{caption}
\usepackage{subcaption}
\usepackage{hyperref}
\usepackage{wasysym}

\definecolor{darkgreen}{rgb}{0,0.5,0}
\makeatletter\def\@fpheader{~}\makeatother
\makeatletter
\newcommand*{\defeq}{\mathrel{\rlap{%
                     \raisebox{0.3ex}{$\m@th\cdot$}}%
                     \raisebox{-0.3ex}{$\m@th\cdot$}}%
                     =} 
\makeatother

\newcommand{\be}{
\begin{eqnarray}
}

\newcommand{\ee}{
\end{eqnarray}
}

\allowdisplaybreaks

\numberwithin{equation}{section}

\usepackage{titlesec}

\titleclass{\subsubsubsection}{straight}[\subsection]

\newcounter{subsubsubsection}[subsubsection]
\renewcommand\thesubsubsubsection{\thesubsubsection.\arabic{subsubsubsection}}

\titleformat{\subsubsubsection}
  {\normalfont\normalsize\bfseries}{\thesubsubsubsection}{1em}{}
\titlespacing*{\subsubsubsection}
{0pt}{3.25ex plus 1ex minus .2ex}{1.5ex plus .2ex}

\makeatletter
\renewcommand\paragraph{\@startsection{paragraph}{5}{\z@}%
  {3.25ex \@plus1ex \@minus.2ex}%
  {-1em}%
  {\normalfont\normalsize\bfseries}}
\renewcommand\subparagraph{\@startsection{subparagraph}{6}{\parindent}%
  {3.25ex \@plus1ex \@minus .2ex}%
  {-1em}%
  {\normalfont\normalsize\bfseries}}
\def\toclevel@subsubsubsection{4}
\def\toclevel@paragraph{5}
\def\toclevel@paragraph{6}
\def\l@subsubsubsection{\@dottedtocline{4}{7em}{4em}}
\def\l@paragraph{\@dottedtocline{5}{10em}{5em}}
\def\l@subparagraph{\@dottedtocline{6}{14em}{6em}}

\def\be{\begin{equation}}
\def\ee{\end{equation}}
\def\ba#1\ea{\begin{align}#1\end{align}}
\def\bg#1\eg{\begin{gather}#1\end{gather}}
\def\bm#1\em{\begin{multline}#1\end{multline}}
\def\bmd#1\emd{\begin{multlined}#1\end{multlined}}

\def\({\left(}
\def\){\right)}
\def\[{\left[}
\def\]{\right]}
\def\<{\langle}
\def\>{\rangle}

\newcommand{\bfig}{\begin{figure}\begin{center}}
\newcommand{\efig}{\end{center}\end{figure}}
\newcommand{\bi}{\begin{itemize}}
\newcommand{\ei}{\end{itemize}}

\newcommand{\Hn}{\mathcal{H}_{\text{p}}}
\newcommand{\Hp}{\mathcal{H}_{\text{np}}}
\newcommand{\mr}[1]{\mathrm{#1}}

\begin{document}
\title{Null states and time evolution in a toy model of black hole dynamics}
\author[a]{Xi Dong,}
\author[a,b]{Maciej Kolanowski,}
\author[a]{Xiaoyi Liu,}
\author[a]{Donald Marolf,}
\author[a,c]{Zhencheng Wang}
\affiliation[a]{Department of Physics, University of California, Santa Barbara, CA 93106, USA}
\affiliation[b]{Institute of Theoretical Physics, Faculty of Physics,
University of Warsaw, Pasteura 5, 02-093 Warsaw, Poland}
\affiliation[c]{Department of Physics, University of Illinois Urbana-Champaign, Urbana, IL 61801, USA}
\emailAdd{xidong@ucsb.edu}
\emailAdd{mkolanowski@ucsb.edu}
\emailAdd{xiaoyiliu@ucsb.edu}
\emailAdd{marolf@ucsb.edu}
\emailAdd{zcwang1@illinois.edu}

\abstract{ Spacetime wormholes can provide non-perturbative contributions to the gravitational path integral that make the actual number of states $e^S$ in a gravitational system much smaller than the number of states $e^{S_{\mr{p}}}$ predicted by perturbative semiclassical effective field theory.  The effects on the physics of the system are naturally profound in contexts in which the perturbative description actively involves $N = O(e^S)$ of the possible $e^{S_{\mr{p}}}$ perturbative states; e.g., in late stages of black hole evaporation.  Such contexts are typically associated with the existence of non-trivial quantum extremal surfaces.
However, by forcing a simple topological gravity model to evolve in time, we find that such effects can also have large impact for $N\ll e^S$ (in which case no quantum extremal surfaces can arise).  In particular, even for small $N$, the insertion of generic operators into the path integral can cause the non-perturbative time evolution to differ dramatically from perturbative expectations.  On the other hand, this discrepancy is small for the special case where the inserted operators are non-trivial only in a subspace of dimension $D \ll e^S$.  We thus study this latter case in detail.  We also discuss potential implications for more realistic gravitational systems.}

%%%%%%%%%%%%%%%%%%%%%%%%%%%%%% MAIN %%%%%%%%%%%%%%%%%%%%%%%%%%%%%%

\maketitle

\section{Introduction}

Perturbative descriptions of quantum gravity are well-known to allow an arbitrarily large number of states inside a black hole with given area $A$.  For example, a large collection of states is readily formed by starting with a black hole of very large area and then letting the black hole decay until it has area $A$.  The tension between this fact and the finite Bekenstein-Hawking entropy forms the core of the so-called black hole information problem.

Recent work on gravitational path integrals and quantum extremal surfaces shows in a rather concrete way how this divergent perturbative result is transformed into a density of states that is finite at the non-perturbative level \cite{Penington:2019npb,Almheiri:2019psf,Penington:2019kki,Almheiri:2019qdq}, and which is in particular given by the generalized entropy $S_{\mr{gen}}$ of an appropriate quantum extremal surface \cite{Engelhardt:2014gca}.  Important components are the inclusion of spacetime wormholes and the choice of a baby-universe superselection sector \cite{Marolf:2020xie}.

At least in simple contexts and with the use of these ingredients, the resulting non-perturbative path integral can be said to define a new inner product on the space of perturbative states.   In particular, certain would-be perturbative states turn out to have vanishing norm with respect to the full non-perturbative inner product \cite{Penington:2019kki,Marolf:2020xie}.  After taking the quotient of the perturbative Hilbert space by such null states, one finds the desired finite density of states; see also \cite{Balasubramanian:2022lnw,Balasubramanian:2022gmo}.

\begin{figure}
    \centering
    \includegraphics[scale=0.8]{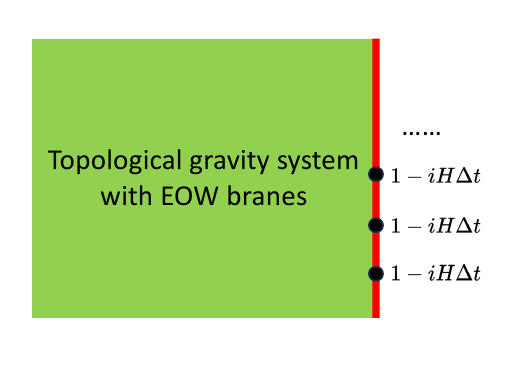}
    \caption{While the original dynamics of the topological model is trivial, we can force the system to evolve in time by inserting a large number of operators $(1-iH\Delta t)$ with small $\Delta t$. }
    \label{fig:sourcedevolution}
\end{figure}

This scenario provides an elegant solution to the state-counting issue.  However, it is clearly of interest to understand any further implications for gravitational physics as well.  Below, we
use the 2d topological model of \cite{Marolf:2020xie} (with end-of-the-world [EOW] branes) to study path integrals that compute time evolution in the presence of repeated boundary sources; see figure \ref{fig:sourcedevolution} above.   It will be convenient to use the language of Euclidean path integrals, so that a real-time evolution is then generated by the insertion of complex sources.  An immediate benefit of this approach will be to show that, in contrast with the concerns expressed in e.g. section 7.3 of \cite{Guo:2021blh}, there is no tension between the presence of null states and unitary evolution.  However, for general such sources, we will nevertheless find the non-perturbative results to differ markedly from perturbative expectations, even in contexts where we study only a small number of initial and final states.

This initial result then motivates us to further investigate a special class of boundary sources for which we expect non-perturbative contributions to be small.  Namely, we consider operators of rank $D$ on the perturbative Hilbert space with $D \ll d$, where $d$ is the dimension of the non-perturbative Hilbert space.  In such cases we find close agreement between the perturbative and non-perturbative path integrals, even when these integrals describe evolution over long times $t$ of order $d^\alpha$ with $\alpha<\frac{1}{2}$.  Furthermore, in this case, we also find there to be a map $\eta$  from the perturbative Hilbert space ${\cal H}_{\mr{p}}$ to the non-perturbative Hilbert space ${\cal H}_{\mr{np}}$ that is approximately equivariant with respect to time evolution.  In other words, if our path integral defines the perturbative time evolution operator $e^{-iHt}$ and the non-perturbative time evolution operator $e^{-i \tilde H t}$, then acting on generic states $|\psi\rangle$ in ${\cal H}_{\mr{p}}$ to good approximation we have
\begin{equation}
e^{-i \tilde H t}\eta  |\psi \rangle  \approx \eta e^{-i H t} |\psi \rangle.
\end{equation}
Here and below we will use the term `generic states' to mean states chosen without knowledge of the baby-universe superselection sector.  Statements about generic states will hold with high probability in the limit where the dimension $d$ of the non-perturbative Hilbert space is large.

We begin with a brief review in section \ref{sec:TopoModel} of  the topological gravity model introduced in \cite{Marolf:2020xie}. Section \ref{sec:HilbertSpaces} then discusses the perturbative and non-perturbative Hilbert spaces ${\cal H}_{\mr{p}}$ and ${\cal H}_{\mr{np}}$ as well as the natural map $\eta$  from ${\cal H}_{\mr{p}}$ to ${\cal H}_{\mr{np}}$. The large discrepancies between perturbative and non-perturbative time evolutions for general sources are studied in section \ref{sec:TimeEvolLR}, while the long-time agreement for repeated sources of small rank is derived in section \ref{sec:TimeEvol}.  We close with further discussion in section \ref{sec:Discussion} emphasizing possible implications for more realistic models of gravity.

\section{The Euclidean gravitational path integral and our topological model}
\label{sec:TopoModel}
The Euclidean gravitational path integral computes correlation functions of observables by integrating over a set of fields $\Phi$ that satisfy given boundary conditions. Such fields are generally taken to include a spacetime metric. For asymptotically Anti-de Sitter (AdS) gravitational theories, the gravitational path integral is usually interpreted as computing partition functions $Z[J]$ for each boundary condition $J$ defined by an asymptotically locally AdS (AlAdS) boundary.   In the context of AdS/CFT, the quantity $Z[J]$ is dual to a corresponding CFT partition function.

We will denote bulk fields $\Phi$ satisfying the boundary condition $J$ by writing $\Phi \sim J$. For boundary conditions on a disconnected boundary manifold with $n$ connected components, and with boundary condition~$J_i$ on the $i$th component, we use the following notation for the Euclidean gravitational path integral:
\begin{equation}
\label{eq:EPI}
    \left\langle Z\left[J_1\right] \cdots Z\left[J_n\right]\right\rangle:=\int_{\Phi \sim J} \mathcal{D} \Phi e^{-S[\Phi]}.
\end{equation}
Note that the description on the right-hand-side depends only on the union of the asymptotic boundaries and is thus manifestly independent of the numbering assigned to the various connected components of the boundary.  As a result, such correlation functions are invariant under permuting the boundary conditions $J_i$; e.g.,
\begin{equation}   \left\langle Z\left[J_1\right]  Z\left[J_2\right] Z\left[J_3\right]\right\rangle =    \left\langle Z\left[J_1\right]  Z\left[J_3\right] Z\left[J_2\right]\right\rangle =    \left\langle Z\left[J_3\right]  Z\left[J_2\right] Z\left[J_1\right]\right\rangle, \  {\rm  etc.}
\end{equation}

A feature of the gravitational path integral is that it generally does not factorize over disconnected boundaries:
\begin{equation}
\label{eq:nofact}
    \langle Z[J_1] Z[J_2] \rangle \neq \langle Z[J_1] \rangle \langle Z[J_2] \rangle.
\end{equation}
 The difference between the two sides arises from the contributions of bulk spacetime wormholes that connect otherwise-separate asymptotic boundaries. This feature is inconsistent with  expectations from AdS/CFT since
CFT partition functions factorize on any disconnected manifold.  However, it is expected that gravitational path integrals that fail to factorize can be instead interpreted as describing averages over a non-trivial ensemble of dual boundary theories; see e.g. \cite{Maldacena:2004rf,Saad:2019lba}.

\begin{figure}
    \centering
    \includegraphics[width=0.3\linewidth]{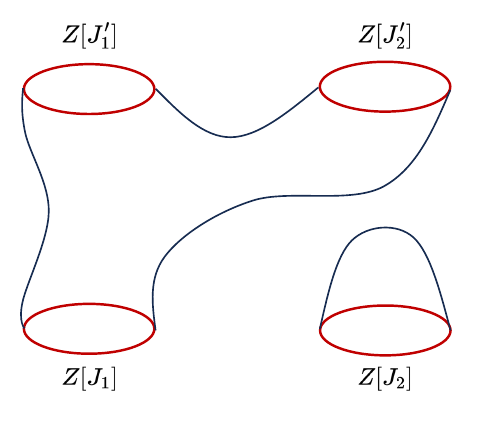}
    \caption{A particular contribution to the path integral that computes $\langle Z[J_1']Z[J_2']| Z[J_1] Z[J_2]\rangle$.}
    \label{fig:ZZZZ}
\end{figure}

When the above ensemble structure arises, it can be seen directly in the bulk description. There it is related to the existence of baby universes \cite{Coleman:1988cy,Giddings:1988cx}.   Such baby universes are necessarily a part of any bulk theory in which the Euclidean gravitational path integral sums over topologies.   The baby universe states
then span a subspace $\mathcal{H}_{\mr{BU}}$ of the full quantum gravity Hilbert space, where
states in $\mathcal{H}_{\mr{BU}}$ can be obtained by slicing open gravitational path integrals on slices that do not intersect any asymptotic boundaries. Our description of the above connections will follow \cite{Marolf:2020xie}, which emphasizes a basis of baby universe states that may be labeled
\begin{equation}
\label{eq:BUbasis}
    |Z[J_1]Z[J_2]\cdots Z[J_n]\rangle,
\end{equation}
where $J_1,\dots, J_n$ again represent a set of closed asymptotic boundaries on which given boundary conditions are satisfied. There is in particular a `no-boundary Hartle-Hawking state' $|\mr{HH}\rangle$ associated with choosing $n=0$. The inner product between two such states is then defined to be
\begin{equation}
\label{eq:ip}
\langle  Z[J'_1]Z[J'_2]\cdots Z[J'_m]   |Z[J_1]Z[J_2]\cdots Z[J_n]\rangle
= \langle  Z[J'_1{}^*]Z[J'_2{}^*]\cdots Z[J'_n{}^*] Z[J_1]Z[J_2]\cdots Z[J_n]\rangle,
\end{equation}
where $*$ denotes an appropriate anti-linear CPT-conjugation operation on the space of boundary conditions; see e.g. Figure \ref{fig:ZZZZ}. One thinks of these boundaries as lying in the asymptotic Euclidean past or future so that space has no boundary at any finite time.
We will assume the above inner product to satisfy reflection positivity:
\begin{equation}
    \|\Psi\|^2:=\langle\Psi | \Psi\rangle \geq 0 \text { for all }|\Psi\rangle=\sum_{i=1}^N c_i\left|Z\left[J_{i, 1}\right] \cdots Z\left[J_{i, m_i}\right]\right\rangle.
\end{equation}
The invariance of correlation functions with respect to permutations of the $J_i$ (described under \eqref{eq:EPI}) then implies the states \eqref{eq:BUbasis} to be similarly invariant under such permutations of their $Z[J_i]$ labels; e.g.    $ |Z[J_1]Z[J_2]\rangle =|Z[J_2]Z[J_1]\rangle.$

A state of the form \eqref{eq:BUbasis} can also be thought of as the result of acting with operators $\widehat{Z[J_i]}$ on the no-boundary Hartle-Hawking state $|\mr{HH}\rangle$.  In fact, we may define the $\widehat{Z[J_i]}$  by the relations:
\begin{equation}
\label{eq:Zops}
    |Z[J_1]Z[J_2]\cdots Z[J_n]\rangle = \widehat{Z[J_1]}\widehat{Z[J_2]}\cdots \widehat{Z[J_n]}|\mr{HH}\rangle.
\end{equation}

Since changing the order of operators does not alter the state, all of the $\widehat {Z[J]}$ operators commute with each other when acting on the dense domain spanned by states of the form \eqref{eq:BUbasis}.  Furthermore, since \eqref{eq:ip} implies that the adjoints satisfy $\widehat{Z[J]}^\dagger =\widehat{Z[J^*]}$ on the given states, these operators also commute with all of the $\widehat{Z[J]}^\dagger$ operators when acting on such states.  As a result, if we assume that such results extend to a larger common domain on which all $\widehat{Z[J]}\pm \widehat{Z[J]}^\dagger$  satisfy the mathematical requirements to be fully self-adjoint (or anti self-adjoint) \cite{RS}\footnote{The reader should be aware that the mathematics literature contains examples where such extensions fail spectacularly; see e.g.  \cite{RS} for an example where symmetric operators that commute on a common invariant dense domain are essentially self-adjoint -- so that they have a unique self-adjoint extension to the full Hilbert space -- but where their self-adjoint extensions nevertheless fail to commute. However, it is far from clear that such failures occur in interesting models of gravitational physics. }, then the $\widehat{Z[J]}$ can all be diagonalized simultaneously. The simultaneous eigenvectors of all such operators are called $\alpha$-states:
\begin{equation}
    \widehat{Z[J]}|\alpha\rangle =Z_\alpha[J]|\alpha\rangle.
\end{equation}
By inserting complete sets of $\alpha$ states, we find an ensemble structure:
\begin{equation}
    \begin{aligned}
\left\langle Z\left[J_1\right] \cdots Z\left[J_n\right]\right\rangle & =\sum_{\alpha_0, \alpha_1, \ldots, \alpha_n}\left\langle\mathrm{HH}|\alpha_0\right\rangle\left\langle\alpha_0\left|\widehat{Z\left[J_1\right]}\right| \alpha_1\right\rangle \cdots\left\langle\alpha_{n-1}\left|\widehat{Z\left[J_n\right]}\right| \alpha_n\right\rangle\left\langle\alpha_n | \mathrm{HH}\right\rangle \\
& =\mathfrak{Z} \sum_\alpha p_\alpha Z_\alpha\left[J_1\right] \cdots Z_\alpha\left[J_n\right],
\end{aligned}
\end{equation}
where $\mathfrak{Z}=\langle \mathrm{HH}|\mathrm{HH} \rangle$ is the norm of the Hartle-Hawking state, and $ p_\alpha=\frac{|\langle\mathrm{HH}|\alpha\rangle|^2}{\langle\mathrm{HH} |\mathrm{HH}\rangle}$ is the probability to measure the state $|\mathrm{HH}\rangle$ to be in the state $|\alpha\rangle$. We also see that correlation functions factorize in $\alpha$ states. For example, when the $\alpha$-states are normalizable we have
\begin{equation}
\left\langle\alpha\left|\widehat{Z\left[J_1\right]} \widehat{Z\left[J_2\right]}\right| \alpha\right\rangle=\left\langle\alpha\left|\widehat{Z\left[J_1\right]}\right| \alpha\right\rangle\left\langle\alpha\left|\widehat{Z\left[J_2\right]}\right| \alpha\right\rangle=Z_\alpha\left[J_1\right] Z_\alpha\left[J_2\right].
\end{equation}

We emphasize that a more standard AdS/CFT scenario, in which the bulk path integral respects factorization, can also be described in the above language.  In that case, one simply finds that there is only one $\alpha$-sector, so that the ensemble becomes trivial.  The interesting consequence is then that there can be only one state in the baby universe Hilbert space $\mathcal H_{\mr{BU}}.$  Assuming that it is not a null state, the Hartle-Hawking no-boundary state  $|\mr{HH}\rangle$ is then non-perturbatively equivalent to the unique $\alpha$-state.

Other sectors of the quantum gravity Hilbert space can be generated by
cutting open the gravitational path integral along slices that intersect asymptotic boundaries on some non-trivial codimension-2 surface  $\Sigma$. The full quantum gravity Hilbert space can be decomposed into sectors labelled by geometry (and perhaps other sources) on such $\Sigma$ so that we may write
\begin{equation}
    \mathcal{H}_{\mr{QG}}=\bigoplus_\Sigma \mathcal{H}_{\Sigma},
\end{equation}
where $\Sigma=\emptyset$ corresponds to $\mathcal{H}_{\mr{BU}}$.  A general Hilbert space $\mathcal{H}_\Sigma$ also admits a decomposition into different $\alpha$ sectors,
\begin{equation}
    \mathcal{H}_\Sigma=\bigoplus_{\alpha} \mathcal{H}_\Sigma^\alpha.
\end{equation}
In particular, the set defined by the labels $\alpha$ is the same for all $\Sigma$; see \cite{Marolf:2020xie} for details.

\subsection{The topological model}
We now review the two-dimensional topological gravity model introduced in \cite{Marolf:2020xie} with $k$ flavors of end-of-the-world-branes (EOW branes). This gives a concrete and solvable model that demonstrates the features mentioned above. The action for the model is given by
\begin{equation}
    S(\mathcal M)=-S_0 \, \chi(\mathcal M)-S_{\partial} \,  n(\mathcal M),
\end{equation}
where the action is evaluated on a compact two-dimensional surface  $\mathcal{M}$  with Euler characteristic $\chi(\mathcal{M})=2-2g(\mathcal{M})-n(\mathcal{M})$ where $g(\mathcal{M})$ is the genus of $\mathcal M$ and $n(\mathcal{M})$ is the number of circular boundaries (including both those that are determined by the boundary conditions $J_i$ and those that arise dynamically from summing over loops of end-of-the-world branes). The theory allows two independent parameters,  $S_\partial$ and $S_0$, but only certain choices satisfy reflection positivity\footnote{The possible choices are $e^{S_{\partial}-S_0} \in \mathbb{N}$ due to the observation of \cite{Marolf:2020xie} that the line segment defines a projection with trace
$e^{S_{\partial}-S_0}$ in one of the $\alpha$-sectors and the observation of
\cite{Colafranceschi:2023urj} that positive-definiteness  of the inner product requires traces of projections to take values in $\mathbb{N} \cup \{+\infty\}$.}. For simplicity, we will use the choice $S_\partial=S_0$ below.

Because we consider a model with EOW branes, the surfaces $\mathcal M$ over which we will sum are allowed to have additional so-called dynamical boundaries, which are boundaries described entirely by EOW branes {\it not} specified by the asymptotic boundary conditions.  Such dynamical boundaries must be labeled by some flavor $I$ of the EOW branes.  Circular $(S^1)$ dynamical boundaries of the same flavor are considered indistinguishable.

The path integral for the model sums over all diffeomorphism classes of two-dimensional oriented surfaces that satisfy given boundary conditions:
\begin{equation}\label{eq:pathIntegral}
    \int_{\Phi \sim J} \mathcal{D} \Phi e^{-S[\Phi]}:=\sum_{\text {Surfaces } \mathcal M \sim J} \mu(\mathcal M) e^{-S[\mathcal M]},
\end{equation}
where $\mu(\mathcal M)=\frac{1}{\prod_g m_{g} ! \prod_{I=1}^k n_I !}$ if $\mathcal{M}$ has $m_g$ connected components of genus $g$ that have only dynamical boundaries, and if $\mathcal M$ has $n_I$ dynamical boundaries associated with EOW-brane flavor $I$. The factor $n_I$ accounts for residual gauge symmetries under the diffeomorphism-invariance described in \cite{Marolf:2020xie}. Note that all boundaries specified by asymptotic boundary conditions are treated as distinguishable. In other words, we do not quotient by diffeomorphisms that relate disconnected components of such boundaries.

\begin{figure}
    \centering
    \includegraphics[width=0.5\textwidth]{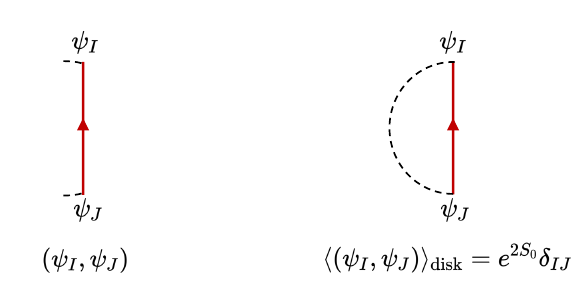}
    \caption{The labelled boundary interval $(\psi_I,\psi_J) $ (left) and the disk contribution to $\langle (\psi_I,\psi_J)\rangle$ (right).}
    \label{fig:psiij}
\end{figure}

Another result of including EOW branes is that there are two kinds of connected asymptotic boundary conditions, which are associated with the following two kinds of operators on $\mathcal H_{\mr{BU}}$:
\begin{itemize}
    \item $\hat Z$, which creates a (unoriented) circular non-dynamical boundary. In the notation introduced earlier we might have written $J=\fullmoon$ and called this operator $\widehat{Z[\fullmoon]}$.
    \item $\widehat{(\psi_J,\psi_I)}$, which creates an oriented interval of non-dynamical boundary running from a label $I$ to a label~$J$.  One may think of such boundaries as describing the creation of an EOW brane with flavor $I$ at the initial endpoint, and then describing the annihilation of a brane with flavor $J$ at the final endpoint.  Surfaces ${\mathcal M}$ in the sum \eqref{eq:pathIntegral} are allowed only if each initial label $I$ is connected to a final label of the same flavor $I$ by a dynamical boundary segment.  In the notation introduced earlier we might have written $J=I\rightarrow K$ and called this operator $\widehat{Z[I \to K]}:=\widehat{(\psi_K,\psi_I)}$; see figure \ref{fig:psiij}.
\end{itemize}
Correlation functions of these operators can be computed from the gravitational path integral; see \cite{Marolf:2020xie} for details.
As described earlier, all of these operators commute, and they also commute with their adjoints. We may thus simultaneously diagonalize them.  Their simultaneous eigenstates are the $\alpha$-states.

After summing over all contributions, one finds that the Hartle-Hawking no-boundary state $|\mr{HH}\rangle$ gives a Poisson distribution for the eigenvalues of $\hat Z$:
\begin{equation}
\label{eq:poisson}
    \mathfrak{Z}^{-1}\langle \hat{Z}^n\rangle=\sum_{d=0}^{\infty} d^n p_d(\lambda), \quad p_d(\lambda)=e^{-\lambda} \frac{\lambda^d}{d !},
\end{equation}
where $d \in \mathbb{N}$, and $\mathfrak{Z}=\langle 1\rangle=e^\lambda$ for $\lambda=\frac{e^{2 S_0}}{1-e^{-2 S_0}}$ is the contribution from summing over all vacuum bubbles. For $d\le k$, we will find below that this Poisson distribution can also be interpreted as an ensemble over $d$-dimensional Hilbert spaces for the one-boundary sector with probabilities $p_d$.

Similar computations show that, within the sector where $Z=d$, the probabilities of the various eigenvalues $(\psi_J,\psi_I)_{d, \tilde \alpha} $ of $\widehat{(\psi_J,\psi_I)}$ are given by a Wishart distribution.  This means that we have\footnote{Note that our normalization differs from that of \cite{Marolf:2020xie}, by an overall factor $1/d$.}
\begin{equation}
\label{eq:wishart}
     \left(\psi_J, \psi_I\right)_{d, \tilde \alpha}=\frac{1}{d}\sum_{a=1}^d \bar{\psi}_J^a \psi_I^a,
\end{equation}
where the $\psi_I^a$ are independent random variables selected from a complex normal (Gaussian) ensemble with unit variance,  and where the symbol $\tilde \alpha$ denotes the collection of $\psi_I^a$ for all $I,a$. Some properties of Wishart distributions are given in appendix \ref{appendix:wishart}. 

A specific $\alpha$-state is defined by choosing a particular instance $(d, \tilde \alpha)$ of both the Poisson and Wishart distributions.   Below, we will simply choose some fixed value of $d$ by hand and then randomly draw the remaining variables $\tilde \alpha$ from the corresponding Wishart distribution.

\section{One-boundary Hilbert spaces and time evolution}
\label{sec:HilbertSpaces}

The discussion in section  \ref{sec:TopoModel} focussed on the baby universe Hilbert space $\mathcal H_{\mr{BU}}$ in order to discuss $\alpha$-sectors in their simplest context.   However, our main interest in the current work will concern the space of one-boundary states; i.e., the space $\mathcal H_\Sigma = \bigoplus_\alpha \mathcal H_\Sigma^\alpha$ for the case where the codimension-2 boundary $\Sigma$ is a single point.  We will use the notation $\mathcal H_{\Sigma_1}^\alpha$ for the superselection sector of the one-boundary Hilbert space labelled by $\alpha$.  This Hilbert space and its perturbative analogue will be described in section \ref{sec:pandnp}. 
Section \ref{sec:form}  clarifies the relation between the perturbative and non-perturbative Hilbert spaces and introduces notation to keep this manifest. Formalism for inserting operators into path integrals of the topological model (and thus inducing time evolution) will be described in section \ref{sec:oite}.  This will set the stage for the analysis of such time evolutions in sections \ref{sec:TimeEvolLR} and \ref{sec:TimeEvol}.

\subsection{The non-perturbative and perturbative Hilbert spaces}
\label{sec:pandnp}

An $\alpha$-sector $\mathcal H_{\Sigma_1}^\alpha$ of the one-boundary Hilbert space is defined by considering all possible boundary conditions $J$ that can be cut into two pieces by using a cut of the form $\Sigma = \Sigma_1$, so that the cut consists only of a single point.  It is clear that any such boundary condition must contain an (oriented) interval $J = I\rightarrow K$ for some $I,K$, which is then cut by the choice of a point $\Sigma_1$ into the pieces $\psi_I := I\rightarrow$ and $\psi_K^* := \rightarrow K$.  The $\psi_I$ define a basis for a linear space $V$ that we may call the space of allowed `Euclidean past' boundary conditions.  Similarly,
the $\psi_K^*$  define a basis for a linear space $V^*$ that we may call the space of allowed `Euclidean future' boundary conditions. We also define an anti-linear map $*: V\rightarrow V^*$ such that  $(\psi_I)^* := \psi^*_I$, and a corresponding map $*: V^*\rightarrow V$ such that  $(\psi_I^*)^* := \psi_I$.

The eigenvalues $(\psi_J,\psi_I)_\alpha$ then define a Hermitian inner product on $V$. Since this inner product was derived taking into account non-perturbative contributions associated with the sum-over-topologies, we will call it the non-perturbative inner product on $V$.  Below, we will use $|I\rangle$ to denote the state $\psi_I \in V$, and we will write the non-perturbative inner product of such states in the form
\begin{equation}
\label{eq:npip}
\langle I|J\rangle_{\mr{np}}:=  \left(\psi_I, \psi_J\right)_{\alpha}.
\end{equation}

Taking the quotient of $V$ by null vectors then yields the Hilbert space $\mathcal H_{\Sigma_1}^\alpha$.  Note that this Hilbert space is automatically complete since $V$ has finite dimension $k$.
It is easy to see from \eqref{eq:wishart} that, with probability one, the right-hand-side of \eqref{eq:wishart} has rank $\min(k,d)$. It is thus clear that $\min(k,d)$ is the dimension of  $\mathcal H_{\Sigma_1}^\alpha$.

In particular, equation \eqref{eq:poisson} defines an ensemble of such Hilbert spaces $\mathcal H_{\Sigma_1}^\alpha$ with different dimensions $d$.  Below, we will imagine that we have already selected a value of $\alpha = (d,\tilde \alpha)$  from this ensemble.  We will then refer to the associated $\mathcal H_{\Sigma_1}^\alpha$ as the non-perturbative Hilbert space $\mathcal H_{\mr{np}}$.

The goal of our study below will be to compare the results of non-perturbative computations with those that would be obtained using a simpler perturbative description that does not include sums-over-topologies.
Such perturbative computations cannot depend on $\alpha$, as the $\alpha$-parameters were already a result of non-perturbative effects from spacetime wormholes.  We thus define the perturbative computation to be performed by taking the baby universes to be in the no-boundary state $|\mr{HH}\rangle$.  With this understanding, we also take the perturbative results to be defined by using only the topology that would give the leading contribution at large $S_0$.  In particular, there is a perturbative Hermitian inner product $\langle I | J\rangle_{\mr{p}}$ defined on $V$ which we take to be given by\footnote{Including this factor of $e^{-2S_0}$ gives a convenient overall normalization, though the results would be equivalent if one chose not to include this factor.} $e^{-2S_0}$ times the disk path integral with asymptotic boundary condition $(\psi_I,\psi_J).$ In particular, we define a conveniently normalized version $\mathcal D$ of the disk amplitude by the relations
\begin{equation}
\label{eq:pertip}
\mathcal D[Z] = 1, \ \ \ {\rm and} \ \ \ \mathcal D \left[(\psi_I,\psi_J) \right] :=e^{-2S_0}\langle (\psi_I,\psi_J)\rangle_{\mathrm{disk}}=\delta_{IJ} =:
  \langle I|J\rangle_{\mr{p}};
\end{equation}
see again Figure \ref{fig:psiij}.
Since this inner product is non-degenerate, it promotes the original linear space of Euclidean past boundary conditions $V$ to a (perturbative) Hilbert space $\mathcal H_{\mr{p}}$.

More generally, when the boundary condition has $n$ connected components we similarly define $\mathcal D$ to be $e^{-2nS_0}$
 times the leading ($n$-disk) contribution to the associated path integral.  With this definition, $\mathcal D$ factorizes over disconnected boundaries; e.g.
\begin{equation}
\label{eq:D2}
\mathcal D \left[(\psi_I,\psi_J) (\psi_K,\psi_L) \right] = \mathcal D \left[(\psi_I,\psi_J) \right] \mathcal D \left[(\psi_K,\psi_L) \right].
\end{equation}

We can think of the non-perturbative inner product as being given by the matrix elements of an $\alpha$-dependent self-adjoint operator $M$ of rank $\min(k,d)$ on the perturbative Hilbert space with
\begin{equation}
\label{eq:npip2}
\langle I|M|J\rangle_{\mr{p}}:= \langle  I| J\rangle_{\mr{np}} =   \left(\psi_I, \psi_J\right)_{\alpha}.
\end{equation}
Using
an
overline to denote the average over the Wishart ensemble \eqref{eq:wishart} with a given fixed value of $d$, we find
\begin{equation}
\label{eq:average}
\overline{\langle I|M|J\rangle_{\mr{p}}}=\delta_{IJ}.
\end{equation}
As a result, the perturbative and non-perturbative inner products coincide \emph{on average},  though they may be very different in particular elements of the ensemble. In particular, we know that the rank of $M$ is $\min(k,d)$ while the rank of the perturbative inner product is $k$.  Nevertheless, since the Wishart distribution is Gaussian, it is clear from \eqref{eq:wishart} that fluctuations in given matrix elements $\langle I|M|J\rangle_{\mr{p}}$ will be small whenever $d$ is large.

\subsection{Formalism and notation}

\label{sec:form}

Recall that the non-perturbative  Hilbert space $\mathcal{H}_{\mr{np}}$  is defined by taking the quotient of $V$ with respect to the space of states of vanishing physical norm; i.e., $\mathcal{H}_{\mr{np}} = V/\mathrm{Ker}\,M$.  Since the perturbative inner product was non-degenerate, we may write $\mathcal{H}_{\mr{p}}=V$.  Calling the above quotient map $\eta$, we may write
\begin{equation}
    \eta: \mathcal{H}_{\mr{p}}\to\mathcal{H}_{\mr{np}}, \ \ |\gamma\rangle \mapsto |\tilde{\gamma}\rangle,
\end{equation}
where $|\tilde \gamma\rangle$ is the equivalence class in $V/\mathrm{Ker}\,M$ of the state $|\gamma\rangle \in V$.
We may alternatively think of $\eta$ as an isomorphism from the quotient $\mathcal{H}_{\mr{p}}/\mathrm{Ker}\,M$  to $\mathcal{H}_{\mr{np}}$.

Note that since the eigenvalues of $M$ are determined by the Wishart distribution, the map $\eta$ is generally not an isometry even when acting on states in $\Hn$ orthogonal to $\mathrm{Ker}\, M$.
 However, it will also be useful to isometrically embed the non-perturbative Hilbert space into $\mathcal{H}_{\mr{p}}$.  In order to do so, let us recall that,
since $M$ is Hermitian and positive semidefinite, there is a well-defined positive square root that we may denote by $X: \mathcal{H}_{\mr{p}} \to\mathcal{H}_{\mr{p}}$.
As a result, if we define a map $\Upsilon:\mathcal{H}_{\mr{np}}\to\mathcal{H}_{\mr{p}}$ by
\begin{equation}
\label{eq:Upsdef}
    \Upsilon|\tilde{\gamma}\rangle :=
    X|\gamma\rangle,
\end{equation}
then $\Upsilon$ is an isometry.  In particular, the perturbative inner product of $\Upsilon |\tilde \gamma \rangle$ and $\Upsilon |\tilde \beta \rangle$ is
$\langle\gamma|\Upsilon^\dagger \Upsilon|\beta\rangle_{\mr{p}}=\langle\gamma|X ^\dagger X|\beta\rangle_{\mr{p}}$, which agrees with $\langle \tilde \gamma | \tilde \beta \rangle_{\mr{np}}$ by \eqref{eq:npip2}.
Here and below, it will often be useful for clarity to decorate some $X$'s with daggers $({}^\dagger$) even though $X$ is self-adjoint (so that $X^\dagger =X$).

It will be useful to note that, as for any isometry, we have the relations $\Upsilon^\dagger \Upsilon = {\mathds 1}_{\mr{p}}$  and $\Upsilon \Upsilon^\dagger  = P_\Upsilon$, where $P_\Upsilon$ is the projection onto the range of $\Upsilon$.  In particular, we have $P_\Upsilon X= X$.
We also comment that it is natural to think of $\Upsilon$ as defined by the $\psi_I^a$ of the Wishart distribution in the sense that there is another (not orthonormal, possibly overcomplete) basis $|\hat a\rangle$ of $\mathcal H_{\mr{np}}$ such that $\Upsilon |\hat a \rangle = \sum_I \psi_I^a |I \rangle$.  Below, we will often use the isometry $\Upsilon$ to identify a state
$|\tilde{\alpha}\rangle$ in the quotient $V/{\rm Ker}\, M$ with $X|\alpha\rangle \in XV \subset V$.  Here $XV$ is the range of the linear operator $X$ acting on the linear space $V$.

Let us now make two further brief observations.  The first is that  $X=\Upsilon\circ\eta$.
The second is that if, for every element of our ensemble, $X$ were of the form $\mathcal{N} P$ for a random projection $P$ and some fixed normalization constant $\mathcal{N}$, then  for $k>d$ the condition \eqref{eq:average} and the fact that $X$ has rank $\min(k,d)$ would imply $\mathcal{N} = \sqrt{k/d}$.   We will show in appendix \ref{sec:RP} that $X$ is indeed well-approximated by an operator of this form in the limit $k \gg d$.

\subsection{Operator insertions and time evolution}
\label{sec:oite}

\begin{figure}
    \centering
    \includegraphics[width=0.6\textwidth]{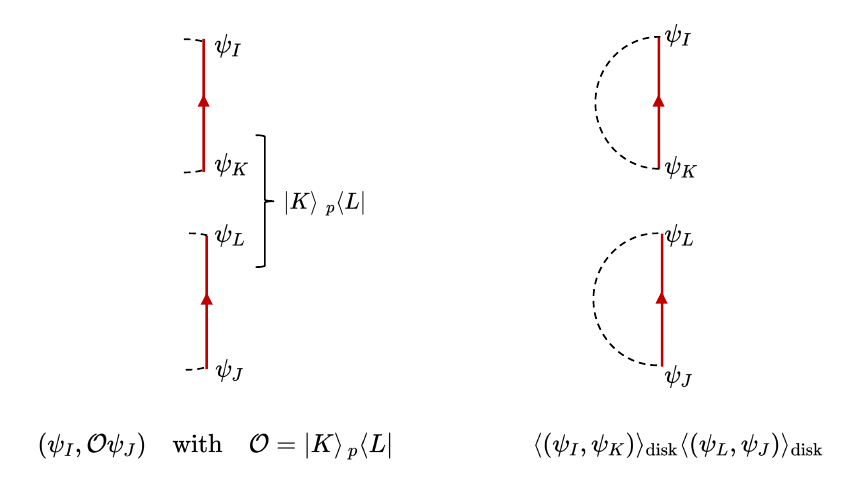}
    \caption{The two labelled intervals $(\psi_I, \psi_K)$ and $(\psi_L, \psi_J)$ (left) that result from the insertion of the operator $|K\rangle\ {}_{\mr{p}}\langle L|$ into $(\psi_I, \psi_K)$, and the disk contribution to the path integral with this boundary condition (right).}
    \label{fig:psiijkl}
\end{figure}

The above inner products are computed by path integrals with a single boundary $(\psi_I,\psi_J)$.  We will be interested in path integrals that describe the further insertion of operators at that boundary.  Recall, however, that the allowed boundary conditions in the model are just unions of the circles ($Z$'s) and labelled intervals $(\psi_I,\psi_J)$.   Suppose then that we wish to insert the perturbative operator $|K\rangle_{\ {\mr{p}}} \langle L|$,
where the notation ${}_{\mr{p}}\langle L|$ indicates that this dual-state acts on vectors in $V$ in the manner defined by the perturbative inner product\footnote{We now see that it would have been better to denote the perturbative inner product of $|I\rangle, |J\rangle \in V$ with a left-subscript  (${}_{\mr{p}}\langle I|J\rangle$) rather than a right-subscript ($\langle I|J\rangle_{\mr{p}}$), though we will continue to use the right-subscript as this notation will be more familiar to many readers.}.  We will then consider boundaries that contain two labelled intervals,
$(\psi_I, \psi_K)$ and $(\psi_L, \psi_J)$ which yields \eqref{eq:D2};
see figure \ref{fig:psiijkl}.
 Of course,
a general operator ${\cal O}$ on $\mathcal{H}_{\mr{p}}$ can be written in the form
\begin{equation}
\label{eq:EOWbasisO}
   \mathcal O=\sum_{K,L}\mathcal O_{K, L}|K\rangle\  {}_{\mr{p}}\langle L|.
\end{equation}
In fact, for any Hilbert space $\mathcal H_{\mr{ext}}$,  we can write a general operator on $\mathcal{H}_{\mr{p}} \otimes \mathcal H_{\mr{ext}}$ in this form by taking the
coefficients $\mathcal O_{K, L}$ to be operators on $\mathcal H_{\mr{ext}}$.
Let us now introduce the symbol $\zeta_\alpha$ to denote the full non-perturbative path integral in a given baby universe $\alpha$-state. Applying $\zeta_\alpha$ to the above boundary condition gives
\begin{eqnarray}
\label{eq:npPIform}
 \zeta_\alpha \left[\sum_{K,L}  \mathcal O_{K, L} (\psi_I, \psi_K)(\psi_L, \psi_J)\right] &=& \sum_{K,L}  \mathcal O_{K, L} \langle I |K\rangle_{\mr{np}}\  \langle L|J\rangle_{\mr{np}} \cr
 &=& \sum_{K,L}  \mathcal O_{K, L} \langle I |M|K\rangle_{\mr{p}}\ \langle L |M| J\rangle_{\mr{p}} \cr  &=& \langle I |X^\dagger\, \left( X \mathcal{O}X^\dagger\right)\, X| J \rangle_{\mr{p}} =
 \langle \tilde I | \Upsilon^{\dagger} \left( X \mathcal{O}X^\dagger\right) \Upsilon | \tilde J \rangle_{\mr{np}}.
\end{eqnarray}
In the last equality we have used the definition \eqref{eq:Upsdef} and the immediate consequence
\begin{eqnarray}
\langle I|X | J \rangle_{\mr{p}} &=&  \langle J|X | I \rangle_{\mr{p}}^*
 =  \langle J| \Upsilon \tilde I \rangle_{\mr{p}}^* \cr
 &=&  \langle \Upsilon^\dagger J| \tilde I \rangle_{\mr{np}}^*
 =  \langle \tilde I | \Upsilon^\dagger J\rangle_{\mr{np}},
\end{eqnarray}
where $|\Upsilon \tilde I\rangle:= \Upsilon |\tilde I\rangle \in XV$ and $|\Upsilon^\dagger J\rangle :=  \Upsilon^\dagger |J\rangle \in V/{\rm Ker}\, M$.
The result \eqref{eq:npPIform} suggests that we should associate any operator ${\cal O}$ on $\Hn$ with a gravitationally dressed operator $\tilde{\cal O}$ on the non-perturbative Hilbert space of the form
\begin{equation}
\label{eq:tildeOdef}
\tilde {\cal O} = \Upsilon^\dagger   X \mathcal{O} X^\dagger \Upsilon.
\end{equation}
Indeed, if we write down a corresponding path integral that, at the disk level, would compute the correlator $\langle I| {\cal O}_1 {\cal O}_2 \dots {\cal O}_n |J \rangle_{\mr{p}}$, then at the non-perturbative level we instead find the result $\langle \tilde I| \tilde {\cal O}_1 \tilde {\cal O}_2 \dots \tilde {\cal O}_n |\tilde J\rangle_{\mr{np}}$. Henceforth, it will be convenient to use $\Upsilon$ to identify $\mathcal H_{\mr{np}}$ with its isometric embedding in $\mathcal H_{\mr{p}}$ as previously advertised.  In doing so, we will write \eqref{eq:tildeOdef} in the simpler form
\begin{equation}
\label{eq:tildeOdefsimp}
\tilde {\cal O} =   X \mathcal{O} X^\dagger.
\end{equation}
In the case where the ${\cal O}_{K,L}$ are operators on some external Hilbert space $\mathcal H_{\mr{ext}}$, it would be more explicit to write \eqref{eq:tildeOdefsimp} in the form
\begin{equation}
\label{eq:tildeOdefsimp2}
\tilde {\cal O} =   \left( X\otimes {\mathds 1}_{\mr{ext}} \right)  \mathcal{O}  \left( X^\dagger\otimes {\mathds 1}_{\mr{ext}} \right),
\end{equation}
where ${\mathds 1}_{\mr{ext}}$ is the identity on $\mathcal H_{\mr{ext}}$.

\begin{figure}[h!]
    \centering
    \includegraphics[scale=0.8]{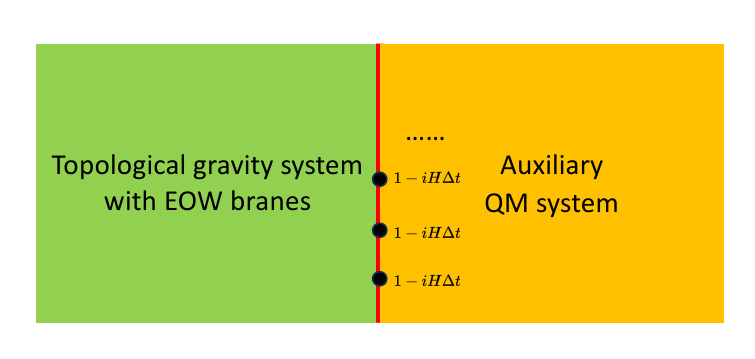}
    \caption{Boundary insertions can be used to couple our topological model to an external non-gravitational `bath' system. }
    \label{fig:coupled}
\end{figure}

Although the original topological model has no evolution (i.e., its Hamiltonian $H_0$ vanishes identically), we can now use this formalism to introduce and study an arbitrary notion of time-evolution for this model.  In particular, let us consider a one-parameter family of operators $H(t)$ on the perturbative one-boundary Hilbert space $\mathcal H_{\mr{p}}$.    Let us then choose times $t_n$ for all $n\in \mathbb{Z}$ such that $t_n \rightarrow \pm \infty$ as $n\rightarrow \pm \infty$ and use the representation
\eqref{eq:EOWbasisO} to define boundary conditions corresponding to the insertion of the sequence of operators $1 - i(t_n - t_{n-1})  H(t_n)$.  In the limit where we increase the density of times $t_n$ such that $t_n-t_{n-1}\rightarrow 0$ for all $n$, the corresponding (rescaled) disk path integral $\mathcal D$ computes matrix elements of the path-ordered-exponential operator ${\cal P}\exp(-\int i H(t) dt)$. It thus implements the time evolution associated with $H(t)$.  As noted above, this can also include the introduction of interactions between our topological model and an external system $\mathcal H_{\mr{ext}}$; see figure \ref{fig:coupled}.

As in figure \ref{fig:psiijkl}, each insertion of an operator of the form \eqref{eq:EOWbasisO} increases the number of connected components of the boundary by one. Since each component computes an inner product, and since the non-perturbative inner product inserts an $X$ (or $X^\dagger$) next to each bra or ket, we
see immediately that the non-perturbative path integral with such insertions describes evolution under the gravitationally dressed Hamiltonian
\begin{equation}
\label{eq:tildeHdef}
\tilde H(t) = XH(t) X^\dagger.
\end{equation}
Furthermore, it is clear from \eqref{eq:tildeHdef} that the dressed Hamiltonian is self-adjoint, so that time evolution on the non-perturbative Hilbert space is unitary.  In contrast, a less careful analysis might instead lead one to believe that one could use interactions with $\mathcal H_{\mr{ext}}$ to evolve our gravitating system from a non-trivial state to a null state (which would be a clear violation of unitarity).  Indeed, one can easily write down scenarios where the {\it perturbative} evolution under  ${\cal P}\exp(-\int i H(t) dt)$ would transform our system from a non-trivial state $|\psi\rangle$ (for which $\eta |\psi\rangle \neq 0$) into a null state annihilated by $\eta$.  However, we also see that non-perturbative corrections will then necessarily modify the dynamics such that the gravitationally dressed evolution ${\cal P}\exp(-\int i \tilde H(t) dt)$ remains unitary.  We will discuss this issue again in section \ref{sec:Discussion}.

\section{Large discrepancies for generic insertions}
\label{sec:TimeEvolLR}

Having introduced both our general formalism and the specific toy model, we now turn to the question of how the perturbative and non-perturbative time-evolutions compare for path integrals of the form described in section \ref{sec:oite}; see again figure \ref{fig:coupled}.   In performing such analysis we will often compute averaged quantities in order to derive results applicable to typical $\alpha$ sectors, but the reader should keep in mind that the dynamics we define will act within each $\alpha$ sector separately.

We consider the case $d<k$ so that the dimensions of $\mathcal H_{\mr{p}}$ and $\mathcal H_{\mr{np}}$ differ, and we will be most interested in the case $d\ll k$.
We may take the initial and final states to be described by the same boundary condition so that we study only expectation values.  This is not actually a restriction since matrix elements of an operator between states $|\phi_1 \rangle$ and $|\phi_2\rangle$ can always be expressed as linear combinations of expectation values of the operator in states of the form $c_1 |\phi_1\rangle + c_2 |\phi_2\rangle$.
For simplicity, we will also take $H(t)$ to be independent of time.  We thus wish to choose $|\psi\rangle \in \mathcal H_{\mr{p}}$ with $|\tilde \psi \rangle :=\eta|\psi\rangle\in \mathcal{H}_{\mr{np}}$ and to compare
\begin{equation}
\label{eq:ptev}
\langle \psi | e^{-iHt} |\psi\rangle_{\mr{p}}
\end{equation}
with
\begin{equation}
\label{eq:nptev}
\langle \tilde \psi | e^{-i\tilde Ht} |\tilde \psi\rangle_{\mr{np}} =   \langle \psi |X^\dagger e^{-i(XHX^\dagger)t}X| \psi\rangle_{\mr{p}}.
\end{equation}

We will shortly find that, even for short times $t$ of order $\sqrt{d/k}$ for $d\ll k$, the evolutions \eqref{eq:ptev} and \eqref{eq:nptev} are starkly different for general choices of $H$ and $|\psi\rangle$.  One way to see this is to compute averages of traces of powers of $\tilde H$ over the Wishart ensemble (again, for a fixed choice of $d$).  Note that, if we choose an orthonormal basis $|\tilde I\rangle$ of $\mathcal H_{\mr{np}}$ for $I=1,\dots, d$, then each state $|\tilde I\rangle$ must be of the form $\eta |I\rangle$ for some $|I\rangle \in V$, and our isometric embedding $\Upsilon$ \eqref{eq:Upsdef} then tells us that the trace of any operator $\tilde {\cal O}$ on $\mathcal H_{\mr{np}}$ can be written in the form
\begin{equation}
\label{eq:nptrace}
{\operatorname{Tr_{\mr{np}}} \tilde O} : = \sum_I \langle \tilde I | \tilde {\cal O} | \tilde I \rangle_{\mr{np}} =
\sum_I \langle  I |X^\dagger   \Upsilon  \tilde {\cal O} \Upsilon^{\dagger} X|I \rangle_{\mr{p}},
\end{equation}
where we recall that $\Upsilon \Upsilon^\dagger = P_\Upsilon$  (where $P_\Upsilon$ is the projection onto the range of $\Upsilon$, which is also the range of $X$) and that $\Upsilon^\dagger \Upsilon  = {\mathds 1}_{\mr{np}}$,
so that $\Upsilon^{\dagger}$ is the left-inverse of $\Upsilon$.
In particular, for $\tilde {\cal O} = \tilde H^n=[\Upsilon^\dagger XHX^\dagger \Upsilon ]^n = \Upsilon^\dagger [XHX]^n\Upsilon$, so that the result \eqref{eq:nptrace} gives
\begin{equation}
\label{eq:nptraceX}
\operatorname{Tr_{np}} (\tilde H^n)  =
\sum_{I=1}^d \langle  I |X^\dagger    [XHX^\dagger]^n  X|I \rangle_{\mr{p}} =\operatorname{Tr_{p}}   ([XHX^\dagger]^n) = \operatorname{Tr_{p}}   ([MH]^n),
\end{equation}
where  $\operatorname{Tr_{np}}$  denotes the trace on the non-perturbative Hilbert space $\mathcal H_{\mr{np}}$, and $\operatorname{Tr_{p}}$ is the trace on the perturbative Hilbert space $\mathcal H_{\mr{p}}$.
The second equality in \eqref{eq:nptraceX} follows from the fact that, since $|\tilde I\rangle_{\mr{np}}$ is an orthonormal basis and $\Upsilon$ is an isometry, the states $\Upsilon |\tilde I\rangle = X|I \rangle $ form an orthonormal basis for the orthogonal complement of the kernel of $X$.

Recall also that  the Wishart ensemble is built from Gaussian variables.   As a result,  so long as one can write the desired expression in terms of $M$ rather than just $X$, ensemble expectation values are straightforward to compute by performing appropriate Wick contractions; see appendix \ref{appendix:wishart}.  In particular, using \eqref{eq:average} immediately yields
\begin{equation}
\label{eq:trtildeH1}
\overline{\operatorname{Tr_{np}} \tilde H} = \operatorname{Tr_p}  H.
\end{equation}
The result \eqref{eq:trtildeH1} is already interesting since $H$ has $k$ eigenvalues while $\tilde H$ has only $d$ eigenvalues.  As a result,  typical eigenvalues of $\tilde H$ must be larger than those of $H$ by a factor of $k/d$.

The result \eqref{eq:trtildeH1} leaves open the possibility that the discrepancy might just be an overall additive shift in the eigenvalues, which would then give only an overall phase in the time evolution.  However, this possibility is easily ruled out by considering the operator $\Delta : = \tilde H - c {\mathds 1}_{\mr{np}}$ with $c=\frac{1}{d}\operatorname{Tr_p}  H$, so that $\overline{\operatorname{Tr_{np}} \Delta }=0.$
Using cyclicity of the trace together with \eqref{appA:two_point} and \eqref{eq:MHM} yields 
\begin{equation}
\label{eq:Delta2}
\overline{\operatorname{Tr_{np}} (\Delta^2)}=\overline{\operatorname{Tr_p}\left(X H X^{\dagger} X H X^{\dagger}-2 c X H X^{\dagger}+c^2 P_\Upsilon \right)} =\operatorname{Tr_p} (H^2).
\end{equation}

Again, the trace on the left hand side of \eqref{eq:Delta2} is performed over a space of dimension $d$, while the trace on the right is over a space of dimension $k$.  As a result, if there is a meaning to discussing typical eigenvalues of $\Delta$, they must have absolute values that are roughly $\sqrt{k / d}$ times the absolute values of typical eigenvalues of $H$.  Since $\sqrt{k/d}$ is large and $\tilde H= c {\mathds 1}_{\mr{np}} +\Delta$, even if we ignore the overall phase the evolution generated by $\tilde H$ will still be more rapid than that generated by the perturbative Hamiltonian $H$. (However, when $\operatorname{Tr_p}  H$ is dominated by its positive eigenvalues it is also true that $\tilde H$ is close to a multiple of the identity in the sense that $c$ is large compared with the magnitudes of the typical eigenvalues of $\Delta$.

We have also computed the next two moments of $\Delta$.  Performing the relevant Wick contractions yields
\begin{equation}
\label{eq:Delta3}
\overline{\operatorname{Tr_{np}} (\Delta^3)}= \frac{d^2+1}{d^2}\operatorname{Tr_p} (H^3),
\end{equation}
and
\begin{equation}
\label{eq:Delta4}
\overline{\operatorname{Tr_{np}} (\Delta^4)}=
 \left( \frac{1}{d}+ \frac{5}{d^2} \right)\operatorname{Tr_p}(H^4) +  \left(\frac{2}{d} + \frac{1}{d^3} \right) \left(\operatorname{Tr_p}(H^2)\right)^2 .
\end{equation}
These results are again consistent with 
typical eigenvalues of $\Delta$ having absolute values that are roughly $\sqrt{k / d}$ times the absolute values of typical eigenvalues of $H$.  The result \eqref{eq:Delta3} also suggests that, at leading order in $k/d$, the eigenvalues of $\Delta$ are distributed evenly on both sides of zero, but that there is a subleading asymmetry dictated by traces of odd powers of $H$. 

Given that the $d\times d$ matrix $\tilde H$ is constructed from the $k\times k$ matrix $H$ by the random compression $\tilde H = XHX^\dagger$, such large discrepancies should not be a surprise.  Indeed, since the ensemble average over $\tilde H$ is invariant under all unitary transformations on $\mathcal H_{\mr{p}}$, and since we expect fluctuations to be small for $k\gg d$, it is natural to expect that $\tilde H$ is proportional to the identity at leading order in $k/d$, and that the remaining part of $\tilde H$ given by $\Delta$ should consist largely of random junk that has little to do with the detailed dynamics of $H$ (though the distribution of eigenvalues will be influenced by the corresponding distribution for $H$). Indeed, if instead the eigenvalues of $\Delta^2$ became sharply peaked in the limit of large $d$ and large $k/d$, then the coefficient of the $\left(\operatorname{Tr_p}(H^2)\right)^2$ term on the right-hand-side of \eqref{eq:Delta4} would approach $1/d$ at large $d$.  The fact that it instead approaches $2/d$ thus supports the idea that the eigenvalues of $\Delta$ are rather random.

Further support for this picture is provided by considering simple special cases with small $d$.  For example, if $d=1$, then $X$ picks out a single random vector in $\mathcal H_{\mr{p}}$ to represent the one non-trivial eigenstate of $\mathcal H_{\mr{np}}$.  Since the direction of this vector is chosen uniformly, up to fluctuations that are small at large $k$ it must consist of roughly equal components along each eigenvector of $H$.  Thus $XHX^\dagger$ is proportional to $P_\Upsilon$ with a coefficient proportional to $\operatorname{Tr_p}(H)$.  Similarly, for $d=2$, the Wishart distribution will build $X$ from a pair of random states.  Such states are nearly orthogonal at large $k$, and both will again have  roughly equal components along each eigenvector of $H$.  So both eigenvalues of $H$ will be roughly equal and will differ only by random fluctuations.  This will then continue to be the case for all $d \ll k$.

\begin{figure}
    \centering
    \includegraphics{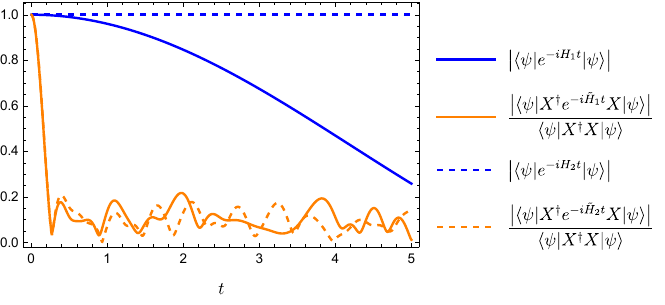}
    \caption{Comparison of two perturbative time evolutions \eqref{eq:ptev} (blue solid and dashed lines) with the corresponding non-perturbative time evolutions \eqref{eq:nptev}  (gold solid and dashed lines) for a fixed random draw from the Wishart ensemble with $k = D= 2000$ and $d=100$ for a randomly chosen state $|\psi\rangle \in \mathcal H_{\mr{p}}$.   The perturbative Hamiltonian $H_1$ was chosen to have a uniformly spaced spectrum  with eigenvalues between $E_{\mathrm{min}}=1$ and $E_{\mathrm{max}}=2$.  In contrast,  $H_2$ is proportional to identity but was normalized to have the same trace as $H_1$. While the details of such plots depend on the random draw from the ensemble, the qualitative features shown above are typical.}
    \label{fig:LargeRank_2}
\end{figure}

As final confirmation of this picture, figure \ref{fig:LargeRank_2} shows numerical results comparing two perturbative time evolutions \eqref{eq:ptev} (blue solid and dashed lines) with the corresponding non-perturbative time evolutions \eqref{eq:nptev}  (gold solid and dashed lines) for a fixed random draw from the Wishart ensemble\footnote{The built-in \textsf{Mathematica} function
   \textsf{WishartMatrixDistribution}   is defined only for  $d\ge k$. In contrast, we are interested in the $k>d$ case. The matrices used to generate our numerical results were thus built using the Mathematical function \textsf{NormalDistribution} and equation (\ref{eq:defM}).}  with $k = D= 2000$ (where $D$ is the rank of $H$) and $d=100$ for a randomly chosen state $|\psi\rangle \in \mathcal H_{\mr{p}}$.   The perturbative Hamiltonian $H_1$ was chosen to have a uniformly spaced spectrum with eigenvalues between $E_{\mathrm{min}}=1$ and $E_{\mathrm{max}}=2$.  In contrast,  $H_2$ is proportional to the identity but was normalized to have the same trace as $H_1$.  While the associated perturbative evolutions are very different, the initial parts of the corresponding non-perturbative time evolutions in figure \ref{fig:LargeRank_2} are nearly identical.  We interpret this as being due to the fact that \eqref{eq:Delta2} implies  typical eigenvalues of $\Delta$ for $H_1$ and $H_2$ to be related by a factor of $\sqrt{\Tr_{\mr{p}}(H_1^2)/\Tr_{\mr{p}}(H_2^2)}\sim 1.018$; i.e., they differ only at the $2\%$ level.  The higher moments will be closely related as well.
   
   Choosing special states $|\psi\rangle$ instead of random states does not appear to improve the situation.  In particular, if $|E_{\mr{max}}\rangle$ is the eigenstate of $H_1$ with highest eigenvalue, taking either $|\psi\rangle = |E_{\mr{max}}\rangle$ or $|\psi\rangle = X|E_{\mr{max}}\rangle$ yields plots that look qualitatively similar to those already shown in figure \ref{fig:LargeRank_2}.

\section{Long-time agreement for small-rank insertions}
\label{sec:TimeEvol}

 Having illustrated the stark difference between perturbative and non-perturbative time evolutions for generic $H$ in section \ref{sec:TimeEvolLR}, it is natural to ask whether there is better agreement for certain choices of $H$. One context where this is to be expected occurs when the rank $D$ of $H$ is small but the dimension $d$ of $\mathcal H_{\mr{np}}$ is large.   At a heuristic level, this follows from the fact that, for a collection of orthonormal states $|I\rangle \in \mathcal H_{\mr{p}}$ (with $I=1,\dots, D$), when $D\ll d$, with high probability one finds the corresponding non-perturbative states $|\tilde I \rangle:=\eta|I\rangle$ to be approximately orthonormal.  As a result, if $H = \sum_{A=1}^D \lambda_A|\phi_A\rangle\langle \phi_A|$, then each $X|\phi_A\rangle$ is an approximate eigenvector of $XHX^\dagger$ with eigenvalue $\lambda_A$.  This means that the dynamics generated by $H$ on the subspace spanned by the $|\phi_A\rangle$ will approximate that generated by $\tilde H$ on the subspace spanned by the images $\eta|\phi_A\rangle$.  Furthermore, since $d\gg D$, a state $|\phi_\perp\rangle$ annihilated by $H$, and thus orthogonal to all $|\phi_A\rangle$, will have $\eta|\phi_\perp\rangle$ approximately orthogonal to all $\eta|\phi_A\rangle$ and will thus be approximately annihilated by $\tilde H$.  Thus the perturbative and non-perturbative path integrals should generate the same dynamics.   The rest of this section will make the above argument precise and will establish quantitative estimates of the associated errors.

We will also choose to make a stronger comparison between the perturbative and non-perturbative time evolutions than in the above sections.  Rather than merely comparing matrix elements of the time evolution operators between corresponding pairs of states, we will now directly compare the final states obtained after time-evolving an initial state.  Specifically, after choosing a state $|\psi\rangle\in\Hn$, we will compare the non-perturbative states
 $e^{i\tilde Ht} \eta |\psi\rangle$ and $\eta e^{iHt} |\psi\rangle$.  Since for typical states $|\phi_1\rangle, |\phi_2\rangle$ the perturbative inner product $\langle \phi_1 | \phi_2\rangle_{\mr{p}}$ is (with high probability) a good approximation to the non-perturbative inner product of
$\eta|\phi_1\rangle$ and $\eta|\phi_2\rangle$, close agreement between $e^{i\tilde Ht} \eta |\psi\rangle$ and $\eta e^{iHt} |\psi\rangle$ will imply (again, with high probability) close agreement between \eqref{eq:ptev} and \eqref{eq:nptev}.  It is in this sense that the comparison below will constitute a more stringent test of agreement than that used in section \ref{sec:TimeEvolLR} above.  We also mention that $\eta e^{iHt} |\psi\rangle$ is the quantity that appears when we can approximate the full set of non-perturbative effects as the insertion of a `black hole final state' (postselection) condition \cite{Horowitz:2003he}.  It is thus clear that this approximation will also fail drastically in the large-rank context that was studied above in section \ref{sec:TimeEvolLR}.

Below, it will often be useful to directly discuss the corresponding operator
\begin{equation}
\label{eq:E1}
E_1(t) := \eta e^{-iHt} = \Upsilon^\dagger X e^{-iHt},
\end{equation}
which enacts perturbative evolution followed by the action of $\eta$, as well as
\begin{equation}
\label{eq:E2}
E_2(t):= e^{-i\tilde{H} t} \eta = \Upsilon^\dagger e^{-iXHX^\dagger t}  X,
\end{equation}
which describes non-perturbative evolution following the action of $\eta$.   In writing \eqref{eq:E1} and \eqref{eq:E2} we have used the relation $\Upsilon^\dagger X = \eta$ which follows from the identities $\Upsilon^\dagger \Upsilon ={\mathds 1}_{\mr{np}}$ and $X = \Upsilon \eta$.

Note that $E_1(t)$ and $E_2(t)$ are both maps from $\Hn$ to $\Hp$. As above, we will generally use the isometry $\Upsilon$ to identify $\mathcal H_{\mr{np}}$ with its image $\Upsilon \mathcal H_{\mr{np}}$ and we will thus omit the factors of $\Upsilon^\dagger$ from $E_1$ and $E_2$.
We will also introduce a notion of `generic' states in $\mathcal H_{\mr{p}}$, meaning that such states are chosen without knowledge of the particular matrix  $M=X^\dagger X$ drawn from the Wishart distribution.
We will see below that for $D\ll d$, even for long times $t$ of order $d^\alpha$ with $\alpha<\frac{1}{2}$, with high probability the operators $E_1(t), E_2(t)$ will agree well when the eigenvalues of $H$ are well-separated and when these operators act on generic states.

\subsection{Eigenvalues and eigenvectors}\label{sec:Preliminaries}
In order to compare \eqref{eq:E1} and \eqref{eq:E2}, it will be useful to first develop some understanding of at least approximate eigenvalues and eigenvectors for the non-perturbative Hamiltonian $\tilde H = X H X^\dagger$.  

Recall from \eqref{eq:average} that the ensemble average of the physical inner product is
\begin{equation}
    \overline{\langle I|X^\dagger X|J\rangle}=\delta_{IJ}.
\end{equation}
We can also use \eqref{appA:two_point} to find
\begin{equation}
    \overline{|\langle I|X^\dagger X|J\rangle|^2}=\overline{M_{IJ}M_{IJ}^*}= \delta_{IJ}+ \frac{1}{d},
\end{equation}
where $I,J$ are not summed. The variance of any given matrix element $\langle I|X^\dagger X|J\rangle$ is thus $1/d$ for all $I,J$.

Denoting the nontrivial eigenvalues of $H$ by $\lambda_A$ and the corresponding orthonormal eigenvectors by $|\phi_A\rangle$ with $A=1,\cdots,D$, we may write the perturbative Hamiltonian $H$ in the form
\begin{equation}\label{eq:expandH}
    H=\sum_{A=1}^D \lambda_A |\phi_A\rangle \langle \phi_A|.
\end{equation}
We will assume that we can write \eqref{eq:expandH} in a form that is independent of $d$ and then study the limit $d\rightarrow \infty$. For the moment, we also assume the eigenvalues to be non-degenerate.  We will consider the degenerate case separately in section \ref{sec:degen}.
The non-perturbative Hamiltonian is correspondingly written
\begin{equation}
\label{eq:expandtildeH}
    \tilde H \equiv XHX^\dagger =\sum_{A=1}^D \lambda_A X|\phi_A\rangle \ {}_{\mr{p}}\langle \phi_A|X^\dagger,
\end{equation}
where the notation ${}_{\mr{p}}\langle  \phi_A|$ again indicates the linear functional on $V = \mathcal H_{\mr{p}}$ that acts on any $|\psi\rangle$ to give the perturbative inner product $\langle \phi_A|\psi\rangle_{\mr{p}}$.

For $D<d$, the $D$ states $X|\phi_A\rangle$ will be linearly independent with probability one (though they will not generally be orthogonal).  This occurs because the states $|\phi_A\rangle$ are linearly independent, so that fine-tuning is required for the action of the map $X$ to render them linearly dependent.  Thus $XHX^\dagger$ has rank $D$ with probability $1$. However, only in the $D \ll d$ limit will the states
$X|\phi_A\rangle$ become approximately orthonormal.  In particular, let us consider applying the Gram-Schmidt orthonormalization procedure to the states  $X|\phi_A\rangle$. So long as $|\phi_A\rangle, |\phi_B\rangle$
are generic states, from \eqref{appA:one_point} and \eqref{appA:two_point}  we know that
\begin{equation}\label{eq:innerProd}
    \langle \phi_A|X^\dagger X|\phi_B\rangle=\delta_{AB} + \epsilon_{AB},
\end{equation}
where
\begin{equation}
    \epsilon_{AB}= |\epsilon_{AB}| e^{i\theta_{AB}},
\end{equation}
$|\epsilon_{AB}|=O(1/\sqrt{d})$,  and  for $A\neq B$ the above $\theta_{AB}$ is a random phase\footnote{However, Hermiticity of $X^\dagger X$ requires $\epsilon_{AA}\in\mathbb{R}$ and thus  $\theta_{AA}=0$ or $\theta_{AA}=\pi$.  More generally, hermiticity of $X^\dagger X$ requires $\epsilon_{AB}=\epsilon_{BA}^*$. }.  As a result, the $D$th vector generated by the Gram-Schmidt procedure will differ from  $X|\phi_D\rangle$ by $D-1$ approximately orthogonal terms of order $\sqrt{1/d}$, so that the magnitude of the difference from $X|\phi_D\rangle$ is of order $\sqrt{\frac{D}{d}}$.   Thus for $D \ll d$ the states $\{X|\phi_A\rangle\}$ are orthonormal with probability $1$.  In
this limit (holding fixed the non-trivial eigenvalues $\lambda_A$),  we see from \eqref{eq:expandtildeH} that the eigenstates of $\tilde H$ become precisely $\{X|\phi_A\rangle\}$ with eigenvalues $\lambda_A$.

We can also expand $\tilde{H}=XHX$ in terms of its own nontrivial  eigenvectors  $|\tilde{\Phi}_A\rangle$ and the corresponding eigenvalues $\tilde{\lambda}_A$. Note that while $|\tilde{\Phi}_A\rangle$ must be $\eta |{\Phi}_A\rangle$ for some state $|\Phi_A\rangle \in V$, this $|\Phi_A\rangle$ will generally not be an eigenvector of the perturbative Hamiltonian $H$.  This is the reason that eigenvectors of $\tilde H$ should not be called $|\tilde \phi_A\rangle.$

As argued above, $\tilde{H}=XHX^\dagger$ has rank $D$ with probability $1$, so  we again take $A=1,\cdots, D$ and write
\begin{equation}
    \label{eq:tildeHinImX}
    \begin{split}
    XHX^\dagger = \Upsilon \tilde{H}\Upsilon^\dagger =&\sum_{A=1}^D \tilde \lambda_A  \Upsilon \frac{|\tilde \Phi_A\rangle \ {}_{\mr{np}}\langle \tilde \Phi_A|}{\langle\tilde\Phi_A|\tilde\Phi_A\rangle_{\mr{np}}} \Upsilon^\dagger
    \\
    =&\sum_{A=1}^D \tilde \lambda_A  \Upsilon \frac{\eta| \Phi_A\rangle \ {}_{\mr{p}}\langle \Phi_A|\eta^\dagger}{\langle \Phi_A|X^\dagger X|\Phi_A\rangle_{\mr{p}}} \Upsilon^\dagger\\
    =&\sum_{A=1}^D \tilde \lambda_A   \frac{X| \Phi_A\rangle \ {}_{\mr{p}}\langle \Phi_A|X^\dagger}{\langle \Phi_A|X^\dagger X|\Phi_A\rangle_{\mr{p}}} ,
    \end{split}
\end{equation}
where in the last step we have used $X= \Upsilon \circ \eta$ and thus $\eta^\dagger \circ \Upsilon^\dagger = X^\dagger$.
The notation ${}_{\mr{p}}\langle \Phi_A|$ again indicates the linear functional on $V = \mathcal H_{\mr{p}}$ that acts on any $|\phi\rangle$ to give the perturbative inner product
$\langle \Phi_A|\phi\rangle_{\mr{p}}$.  Note that we allow arbitrary normalizations for the $|\tilde \Phi_A\rangle$ because we would like to approximate  $\Upsilon |\tilde\Phi_A\rangle$  by $X|\phi_A\rangle = \Upsilon \eta |\phi_A\rangle$.
We also define $\tilde P_0$ to be the projection operator onto the kernel of $\tilde H$.

We now explore the details of how $\tilde \lambda_A, |\tilde\Phi_A\rangle$ approach $\lambda_A, X|\phi_A\rangle$ as $d \to \infty$.
In order to do so, it is useful to note that while states in the kernel of $H$ are orthogonal to those in the range of $H$, this will no longer be true of the images obtained by acting on the above states with the operator $X$.    In particular, for an arbitrary  normalized state $|\phi_0\rangle$,  if the projection map to the kernel of $H$ is $P_0$, we find $\langle\phi_A|X^\dagger XP_0|\phi_0\rangle_{\mr{p}} =O(1/\sqrt{d})$.  

To  avoid writing out this matrix element repeatedly below, we define the symbol  $\varepsilon_A:= \langle\phi_A|X^\dagger XP_0|\phi_0\rangle_{\mr{p}} =O(1/\sqrt{d})$. Furthermore,
from \eqref{appA:one_point} and \eqref{appA:two_point}  we have
\begin{equation}
    \overline{\langle \phi_A|X^\dagger XP_0|\phi_0\rangle_{\mr{p}}}=0,\quad \text{and}\quad \overline{|\langle \phi_A|X^\dagger XP_0|\phi_0\rangle_{\mr{p}}|^2}=\frac{||P_0|\phi_0\rangle||_{\mr{p}}^2}{d}.
\end{equation}
As we will see, the small parameters $\epsilon_{AB}$ and $\varepsilon_A$ will control the discrepancy between the action of $E_1(t)$ and $E_2(t)$ on a generic state. We will compute this discrepancy to leading order in these parameters.

We begin by studying the rate at which each non-perturbative eigenvalue $\{\tilde\lambda_A\}$  converges to  $\{\lambda_A\}$. In particular, we use \eqref{eq:tildeHdef}, \eqref{eq:expandH}, and \eqref{eq:innerProd} to write
\begin{equation}
\label{eq:tHXphi}
    \tilde H X|\phi_B\rangle =X H X^\dagger X|\phi_B\rangle =\sum_A\lambda_A X|\phi_A\rangle\langle\phi_A|X^\dagger X|\phi_B\rangle_{\mr{p}}= \lambda_B X|\phi_B\rangle  +O\left(\sqrt{\frac{D}{d}}\right).
\end{equation}
To see that the error term is of order $\sqrt{\frac{D}{d}}$, one may calculate the norm of the difference between the left side and the explicit term on the right. This  yields
\begin{equation}  
    \overline{\left\lVert \left( \tilde{H}X - \lambda_B X \right)|\phi_B \rangle \right\rVert ^2} = \frac{\operatorname{Tr_p} (H^2)}{d} + \frac{\lambda_B^2 + (\operatorname{Tr_p} \, H)^2}{d^2}.
\end{equation}
Since $\operatorname{Tr_p} (H^2), \operatorname{Tr_p} (H) = O(D)$, it follows that the norm squared is of order $\frac{D}{d}$, hence the error term in \eqref{eq:tHXphi}.

While this result already shows $\tilde \lambda_A \rightarrow \lambda_A$ and $X|\phi_A\rangle \rightarrow |\tilde \Phi_A\rangle$ as $d/D\rightarrow \infty$, the error terms can be characterized more explicitly by introducing the quantities
\begin{equation}
    \delta\lambda_A\equiv \tilde\lambda_A-\lambda_A \quad {\rm and} \quad |\Delta_A\rangle\equiv |\tilde \Phi_A \rangle - X|\phi_A \rangle.
\end{equation}
The normalization of $|\tilde\Phi_A\rangle$ will be chosen to set $\langle \Delta_A|X|\phi_A\rangle_{\mr{p}}=0$. We then have $\delta \lambda_A= O(1/\sqrt{d}), |\Delta_A\rangle = O(\sqrt{\frac{D}{d}})$, where the notation $|\Delta_A\rangle = O(\sqrt{\frac{D}{d}})$ means $\left\lVert |\Delta_A \rangle \right\rVert_{\mr{p}} = O(\sqrt{\frac{D}{d}})$; i.e., $\langle \Delta_A | \Delta_A \rangle_{\mr{p}} = O(D/d)$. Thus we find
\begin{equation}
\begin{split}
\tilde \lambda_A |\tilde \Phi_A\rangle &=
\tilde H |\tilde\Phi_A\rangle =XHX^\dagger |\tilde\Phi_A\rangle\\ &=
\sum_B \lambda_B X |\phi_B\rangle\  {}_{\mr{p}}\langle \phi_B | X^\dagger \left(X | \phi_A \rangle+|\Delta_A\rangle\right)\\
&= \lambda_A X |\phi_A\rangle +  \sum_B \lambda_B X |\phi_B \rangle \epsilon_{BA} + \sum_B \lambda_B X |\phi_B \rangle \langle \phi_B | X^\dagger |\Delta_A \rangle_{\mr{p}},
\end{split}
\end{equation}
where in the last step we have again used equation (\ref{eq:innerProd}).
At the same time, we have
\begin{equation}
\tilde \lambda_A |\tilde \Phi_A\rangle = (\lambda_A+\delta\lambda_A)(X|\phi_A\rangle+|\Delta_A\rangle)=\lambda_A X |\phi_A\rangle + \delta \lambda_A X |\phi_A\rangle + \lambda_A | \Delta_A \rangle+O(\sqrt{D}/d).
\end{equation}
Putting these together yields
\begin{equation}\label{eq:almost}
\delta \lambda_A X |\phi_A\rangle + \lambda_A | \Delta_A \rangle
=   \sum_B \lambda_B X |\phi_B \rangle \epsilon_{BA}+ \sum_B \lambda_B X |\phi_B \rangle \langle \phi_B | X^\dagger |\Delta_A \rangle_{\mr{p}} + O(\sqrt{D}/d).
\end{equation}
Taking the inner product with $X|\phi_A \rangle$ and keeping only first-order corrections we obtain
\begin{equation}
\label{eq:deltalambdai}
\delta \lambda_A = \lambda_A \epsilon_{AA}+O(D/d).
\end{equation}
In computing the above error term, we have noted that the range of $XHX^\dagger$ is spanned by the states $X|\phi_B\rangle$ so that, since we defined  $|\Delta_A\rangle$ to be orthogonal to $X|\phi_A\rangle$, $|\Delta_A\rangle$ is in fact a superposition of the $D-1$ states $X|\phi_B\rangle$ for $B\neq A$.  We have then made the natural assumption that the orders of 
each matrix element $\langle \phi_B | X^\dagger |\Delta_A \rangle_{\mr{p}}$ in both $d$ and $D$ are independent of  $B$, and thus that they are of order $(D^0/\sqrt{d})$. This assumption will be justified below. We also used the following relation
\begin{equation}
    \sum_B \lambda_B \langle \phi_A|X^\dagger X |\phi_B \rangle \epsilon_{BA} = \lambda_A \epsilon_{AA} + \sum_B \lambda_B |\epsilon_{AB}|^2,
\end{equation}
which follows from \eqref{eq:innerProd}.
If we now insert \eqref{eq:deltalambdai} into equation (\ref{eq:almost}), take the inner product with $X|\phi_B\rangle$,  and keep only first-order corrections we find
\begin{equation}
\label{eq:delxphi}
    \langle \Delta_A|X|\phi_B\rangle_{\mr{p}} = \frac{\lambda_B}{\lambda_A-\lambda_B} \epsilon_{AB}(1-\delta_{AB})+O(D/d),
\end{equation}
whence (\ref{eq:almost}) yields
\begin{equation}
\label{eq:Dellast}
    |\Delta_A\rangle = \sum_{B:B\neq A}\frac{\lambda_B}{\lambda_A-\lambda_B}\epsilon_{BA} X|\phi_B\rangle+O(D^{3/2}/d).
\end{equation}
In particular, the leading term in \eqref{eq:delxphi} is sufficient to justify the above assumption stated below \eqref{eq:deltalambdai}.  This establishes the error bound stated in \eqref{eq:deltalambdai} and thus also in \eqref{eq:Dellast}.  

To leading order in the small parameters $\epsilon_{AB}$ and $\varepsilon_A$ we then obtain
\begin{equation}
\langle\tilde\Phi_A|X|\phi_B\rangle_{\mr{p}}=(\ {}_{\mr{p}}\langle\phi_A|X^\dagger+{}_{\mr{p}}\langle\Delta_A|)X|\phi_B\rangle=\delta_{AB}+\epsilon_{AB}+\frac{\lambda_B}{\lambda_A-\lambda_B}\epsilon_{AB}(1-\delta_{AB})+O(D/d).
\end{equation}
Similarly, for any fixed state $|\phi_0\rangle$, we have
\begin{equation}
    \langle \Delta_A|X P_0|\phi_0\rangle_{\mr{p}}=O(D^{3/2}/d).
\end{equation}

We can also use the above results to extract leading order corrections for the projection $\tilde P_0$ onto the kernel of $\tilde H$.
 In particular, we find
\begin{equation}
\label{eq:projection_kernel}
\begin{split}
\tilde{P}_0&:=P_\Upsilon-\sum_A \frac{X|{\Phi}_A\rangle\ {}_{\mr{p}}\langle{\Phi}_A|X^\dagger}{\langle\tilde{\Phi}_A|\tilde{\Phi}_A\rangle_{\mr{np}}}\\&= P_\Upsilon-\sum_A X\left[\left(1-\epsilon_{AA}\right)|\phi_A\rangle\ {}_{\mr{p}}\langle\phi_A|+\sum_{B:B\neq A}  \frac{\lambda_B}{\lambda_A-\lambda_B}(\epsilon_{BA} |\phi_B\rangle\ {}_{\mr{p}} \langle\phi_A|+\epsilon_{AB}|\phi_A\rangle\ {}_{\mr{p}}\langle\phi_B|)\right]X^\dagger + O\left(\frac{D^{3/2}}{d} \right).
\end{split}
\end{equation}
In contrast, the projection onto the kernel of $H$ is $P_0={\mathds 1}_{\mr{p}}-\sum_A |\phi_A\rangle\ {}_{\mr{p}}\langle \phi_A|$.

\subsection{Evolutions of a generic state}
\label{sec:longtimegeneric}
With the above results in hand, we can now compare the actions of $E_1(t)$, $E_2(t)$ on  a generic state $|\phi\rangle$. We remind the reader that we use the term `generic' to refer to states that are selected independently of the random variables in the Wishart distribution. Since we assumed the form of \eqref{eq:expandH} to be fixed (in the sense that it is independent of $d$), we have effectively required $D$ to be of order one as $d\rightarrow \infty$.  We will thus cease to keep track of factors of $D$ in the error terms below.

Let us write our state in the form
\begin{equation}
|\phi\rangle=P_0|\phi\rangle+\sum_{A=1}^D c_A|\phi_A\rangle.
\end{equation}
Perturbative evolution followed by acting with $X$ thus yields
\begin{equation}
\label{eq:E1eval}
    E_1(t)|\phi\rangle=X e^{-i H t}|\phi\rangle=X P_0|\phi\rangle+\sum_A c_A e^{-i \lambda_A t} X|\phi_A\rangle .
\end{equation}
On the other hand, acting with $X$ and then evolving non-perturbatively gives 
\begin{equation}
\label{eq:E2eval}
\begin{aligned}
    E_2(t)|\phi\rangle&=e^{-i \tilde{H} t} X|\phi\rangle=\tilde{P}_0 X|\phi\rangle+\sum_A e^{-i \tilde \lambda_A t} \frac{X| \Phi_A\rangle \ {}_{\mr{p}}\langle\Phi_A|X^\dagger}{\langle\tilde\Phi_A|\tilde\Phi_A\rangle_{\mr{np}}} X| \phi\rangle\\
    &=\tilde P_0 X|\phi\rangle +\sum_{A} e^{-i\tilde \lambda_A t} \frac{X|\Phi_A\rangle \ {}_{\mr{p}}\langle \Phi_A|X^\dagger}{\langle\tilde\Phi_A|\tilde\Phi_A\rangle_{\mr{np}}} X P_0|\phi\rangle+ \sum_{AB} c_B e^{-i\tilde \lambda_A t} \frac{X| \Phi_A\rangle \ {}_{\mr{p}}\langle\Phi_A|X^\dagger}{\langle\tilde\Phi_A|\tilde\Phi_A\rangle_{\mr{np}}} X |\phi_B\rangle\\
    &=XP_0|\phi\rangle+\sum_A c_A X|\phi_A\rangle - \sum_A \frac{X|{\Phi}_A\rangle\ {}_{\mr{p}}\langle{\Phi}_A|X^\dagger}{\langle\tilde{\Phi}_A|\tilde{\Phi}_A\rangle_{\mr{np}}} X\left(P_0|\phi\rangle +  \sum_B c_B |\phi_B\rangle\right)\\
    &\quad +\sum_A \frac{X|{\Phi}_A\rangle\ {}_{\mr{p}}\langle{\Phi}_A|X^\dagger}{\langle\tilde{\Phi}_A|\tilde{\Phi}_A\rangle_{\mr{np}}} e^{-i\tilde{\lambda}_A t} \left( XP_0|\phi\rangle +\sum_B c_B X|\phi_B\rangle \right).
\end{aligned}
\end{equation}
We can now use the results of section \ref{sec:Preliminaries} to further evaluate the final line of \eqref{eq:E2eval}.  This yields
\begin{equation}
\begin{split}
        &\sum_A e^{-i\tilde{\lambda}_A t} \Bigl(  \varepsilon_A X|\phi_A\rangle +c_A(1-\epsilon_{AA})X|\phi_A\rangle +\sum_B \epsilon_{AB} c_B X|\phi_A\rangle  \cr +&\sum_{B:B\neq A} \frac{\lambda_B}{\lambda_A-\lambda_B} \Bigl[c_A \epsilon_{BA} X|\phi_B\rangle +c_B\epsilon_{AB}X|\phi_A\rangle\Bigr]\Bigr)  + O\left(\frac{1}{d}\right)\\
        =&\sum_A \left(e^{-i\tilde{\lambda}_A t} ( c_A + \varepsilon_A )  +\sum_{B:B\neq A} c_B \frac{\lambda_A }{\lambda_A-\lambda_B}\epsilon_{AB}(e^{-i\tilde{\lambda}_A t}-e^{-i\tilde{\lambda}_B t})\right) X|\phi_A \rangle + O\left(\frac{1}{d}\right).
\end{split}
\end{equation}

We can then similarly expand the other terms in \eqref{eq:E2eval}. For example, using \eqref{eq:projection_kernel} we find
\begin{equation}
    \frac{X|{\Phi}_A\rangle\ {}_{\mr{p}}\langle {\Phi}_A|X^\dagger}{\langle\tilde{\Phi}_A|\tilde{\Phi}_A\rangle_{\mr{np}}} = X\left[\left(1-\epsilon_{AA}\right)|\phi_A\rangle\ {}_{\mr{p}}\langle\phi_A|+\sum_{B:B\neq A}  \frac{\lambda_B}{\lambda_A-\lambda_B}(\epsilon_{BA} |\phi_B\rangle\ {}_{\mr{p}} \langle\phi_A|+\epsilon_{AB}|\phi_A\rangle\ {}_{\mr{p}}\langle\phi_B|)\right]X^\dagger + O\left(\frac{1}{d}\right).
\end{equation}
From this, it follows that
\begin{subequations}
    \begin{equation}
        \frac{X|{\Phi}_A\rangle\ {}_{\mr{p}}\langle{\Phi}_A|X^\dagger}{\langle\tilde{\Phi}_A|\tilde{\Phi}_A\rangle_{\mr{np}}} X P_0 |\phi \rangle =   \varepsilon_A X |\phi_A \rangle + O\left(\frac{1}{d}\right)
    \end{equation}
    and
    \begin{equation}
        \frac{X|{\Phi}_A\rangle\ {}_{\mr{p}}\langle{\Phi}_A|X^\dagger}{\langle\tilde{\Phi}_A|\tilde{\Phi}_A\rangle_{\mr{np}}} X |\phi_C \rangle = \delta_{AC}X|\phi_A \rangle + (1- \delta_{AC}) \epsilon_{AC} \frac{\lambda_A}{\lambda_A - \lambda_C} X |\phi_A \rangle + \sum_{B:B \neq A} \frac{\lambda_B}{\lambda_A - \lambda_B} \epsilon_{BA} \delta_{AC} X |\phi_B\rangle + O\left(\frac{1}{d}\right).
    \end{equation}
\end{subequations}
Substituting these results into \eqref{eq:E2eval} we obtain
\begin{equation}
\label{eq:Ediff2}
(E_2(t)-E_1(t))|\phi\rangle = \sum_{A=1}^D \delta_A X|\phi_A\rangle,
\end{equation}
where
\begin{equation}
\label{eq:deltai2}
    \delta_A=c_A \left(e^{-i\tilde \lambda_A t}-e^{-i \lambda_A t}\right)+\left(e^{-i \tilde\lambda_A t}-1\right) \epsilon_A+\sum_{B:B\neq A}c_B \frac{\lambda_A}{\lambda_A-\lambda_B} \epsilon_{AB} (e^{-i\tilde \lambda_A t} - e^{-i\tilde \lambda_B t}).
\end{equation}

From \eqref{eq:deltalambdai} we have $\tilde \lambda_A-\lambda_A = {\cal O}(1/\sqrt d)$,  so that the first term in \eqref{eq:deltai2} may be rewritten in the form
\begin{equation}
c_A e^{-i\tilde \lambda_A t}\left(1 - e^{i(\tilde \lambda_A-\lambda_A) t}\right)
= c_A O\left(t [\lambda_A-\tilde \lambda_A] \right) = c_A O\left(\frac{t}{\sqrt{d}}\right).
\end{equation}
Thus for $t \ll \sqrt{d}$ this term remains very small.

The second term in \eqref{eq:deltai2} is a bounded function of time with amplitude of order  $\frac{1}{\sqrt{d}}$.  However, the third term is more complicated.  It clearly becomes large for $\lambda_A-\lambda_B$ small, so let us assume these differences to be $O(1)$.  The term then involves a sum over a set of bounded functions of time with coefficients of order $c_B/\sqrt{d}$.  The functions of time are small when $t$ is small, after which they will generically contribute random phases.  Thus we see that this sum will again be of order $\frac{\sqrt{\sum_B |c_B|^2}}{\sqrt{d}} < 1/\sqrt{d}$ so that we find
\begin{equation}
\delta_A =  O(1/\sqrt{d}) +  c_A O(t/\sqrt{d}) \ \ \ {\rm for} \ \ \ t < O(\sqrt{d}).
\end{equation}
As a result, for any generic state $|\phi\rangle$ we find 
\begin{equation}
\label{eq:Ediff}
(E_2(t)-E_1(t))|\phi\rangle = O(\sqrt{1/d}) +  O(t/\sqrt{d}) \ \ \ {\rm for} \ \ \ t < O(\sqrt{d}).
\end{equation}

\begin{figure}[h!]
    \centering
    \includegraphics[width=0.9\linewidth]{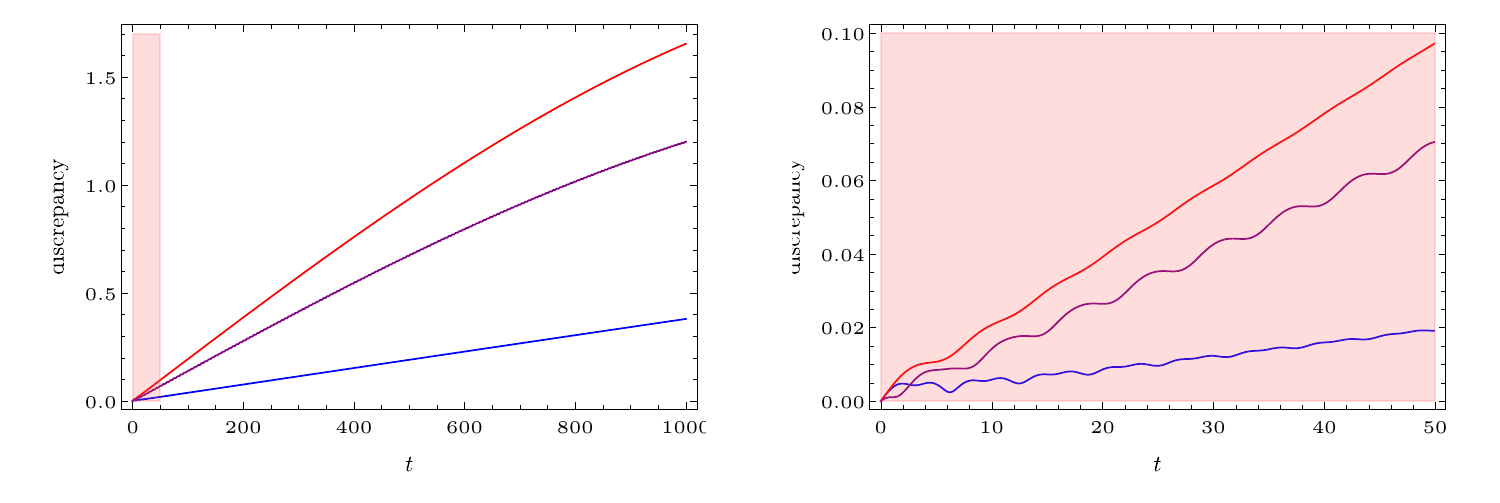}
    \caption{The discrepancy \eqref{eq:discr} for a small-rank Hamiltonian ($k=2d,d=10^6,\sqrt{d}=1000,D=4$) with eigenvalues $\lambda_1=0.5 ,\lambda_2=1.5 , \lambda_3=2.5, \lambda_4=3.5$ and  a random non-perturbative inner product drawn from the Wishart distribution \eqref{eq:wishart}.  Results are shown  for the second perturbative eigenstate $|\phi_2\rangle$ (lower curve in blue), the third perturbative eigenstate $|\phi_3\rangle$ (upper curve in red), and the linear combination $\frac{1}{\sqrt{2}}\left(|\phi_2\rangle + |\phi_3\rangle \right)$ of these eigenstates  (middle curve in purple).  As emphasized in the right panel, the discrepancy is small for $t \ll \sqrt{d}$, but it becomes of order 1 when $t \sim \sqrt{d}$ as shown on the left.  There are $O(1)$ variations between random draws from the Wishart enesmble, though the qualitative form shown is common to many such draws.   }
    \label{fig:smallrank2}
\end{figure}
\begin{figure}[h!]
    \centering
    \includegraphics[width=\textwidth]{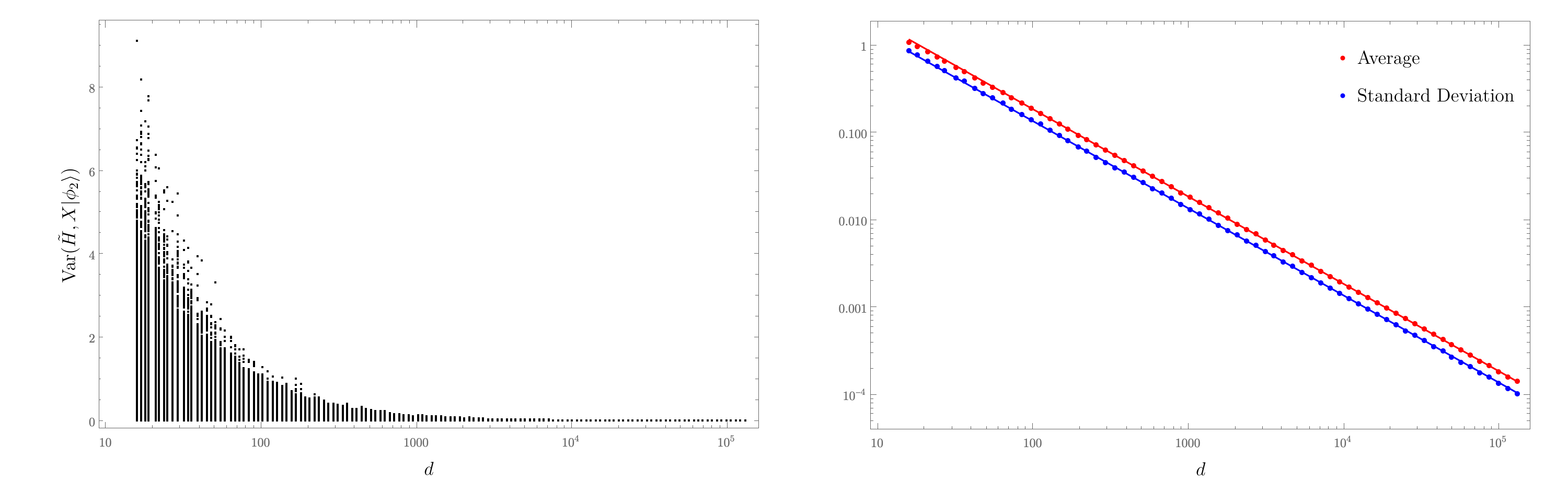}
    \caption{For each $d$, $10\,000$ random draws from the Wishart ensemble for $k=2d$ were used to compute the quantity $\mathrm{var}(\tilde H,X\phi_2):= \frac{\langle\phi_2|X^\dagger XH X^\dagger XHX^\dagger X |\phi_2\rangle}{\langle\phi_2|X^\dagger X|\phi_2\rangle}-\left(\frac{\langle\phi_2|X^\dagger XHX^\dagger X|\phi_2\rangle}{\langle\phi_2|X^\dagger X|\phi_2\rangle}\right)^2$ for the $D=4$ Hamiltonian with $\lambda_1=0.5, \lambda_2=1.5, \lambda_3=2.5, \lambda_4=3.5$.  For each draw, this quantity encodes the variance (squared uncertainty) of $XHX$ in the state $X|\phi_2\rangle$.  Since $X|\phi_2\rangle$ becomes an eigenstate at large $d$, $var(H,X\phi_2)$ should tend to $0$ with probability one as $d\to \infty$ (with an error of order $1/d$). In the left panel, the data is shown in black dots, which indeed asymptotes to $0$ as $d\to\infty$. In the right panel, the red and blue dots show the average value and standard deviation of the data in the left panel for each $d$. The red and blue solid lines are the best fit of the data for $d\ge 512$ to a curve of the form $A\cdot d^{-B}$. We find $B\sim 1.000$ and $B\sim 0.998$ for the average and standard deviation, respectively. }
    \label{fig:diff}
\end{figure}
\begin{figure}[h!]
    \centering
    \includegraphics[width=\textwidth]{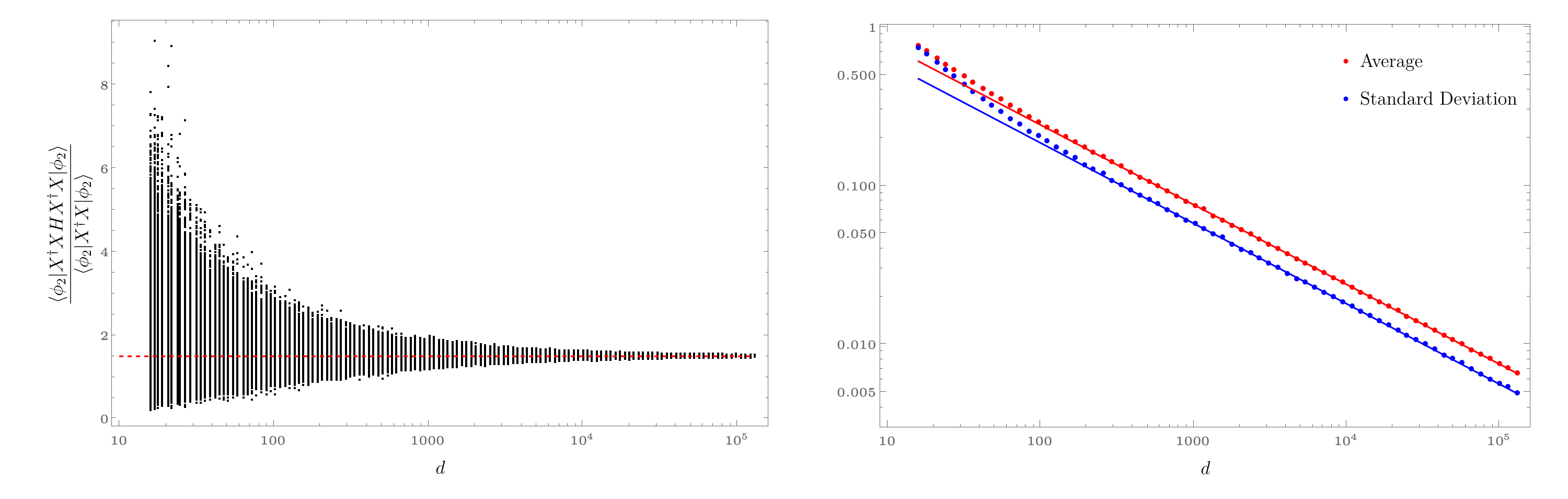}
    \caption{For each value of $d$ shown, 
    the $10,000$ random draws from the Wishart ensemble from figure \ref{fig:diff} are used to compute $\frac{\langle \phi_2|X^\dagger XHX^\dagger X|\phi_2\rangle}{\langle\phi_2|X^\dagger X|\phi_2\rangle}$ for the same $H$ as in figure \ref{fig:diff}. In the left panel, the data is shown in black dots and the dotted red line indicates the eigenvalue $\lambda_2=1.5$ of $H$ to which the black dots should converge at large $d$. In the right panel, the red and blue dots show the average value and standard deviation of the  absolute value of the residuals from the left panel for each $d$. The red and blue solid lines are the best fit of the data for $d\ge 512$ to a curve of the form $A\cdot d^{-B}$. We find $B\sim 0.503$  for the average and $B\sim 0.507$ for the standard deviation,.}
    \label{fig:eigenvalues}
\end{figure}

A numerical example is shown in figure \ref{fig:smallrank2}.  For this purpose we define a real-valued measure of the discrepancy between $E_1(t)$ and $E_2(t)$ acting on the given state $|\phi\rangle:$
\begin{equation}
\label{eq:discr}
    \text{discrepancy}\equiv\frac{\|(E_1(t)-E_2(t))|\phi\rangle\|}{\|X|\phi\rangle\|}.
\end{equation}
We also include figures \ref{fig:diff} and \ref{fig:eigenvalues} which show numerical studies of the convergence of the eigenvectors  of $XHX^\dagger$ to $X|\phi_A\rangle$ and the associated eigenvalues to $\lambda_A$. The results support the claims that $|\Delta_A\rangle$ and $\delta \lambda$ are of order $1/\sqrt{d}$.

In fact, the discrepancy is even smaller than that described above when $|\phi\rangle$  is a randomly chosen state in $\mathcal H_{\mr{p}}$ as then $\sum_{A=1}^D |c_A|^2  = O\left(\frac{D}{k}\right)$.  However, in that case the improvement is due to the fact that, because $H$ was chosen to be of small rank,  most of the state simply does not evolve at all under either $E_1$ or $E_2$. Such states may thus be of less interest.

As described in section \ref{sec:oite}, our formalism also allows us to discuss perturbative Hamiltonians $H$ that couple our topological gravity model to a non-gravitational `bath' system with Hilbert space ${\mathcal H}_{\mr{ext}}$.  We can apply the above analysis directly to this case when there is a $D$-dimensional subspace ${\mathcal H}_{\mr{dynamical}} \subset {\mathcal H}_{\mr{p}}$ such that $H$ acts nontrivially only on 
 ${\mathcal H}_{\mr{dynamical}}\otimes {\mathcal H}_{\mr{ext}}\subset {\mathcal H}_{\mr{np}}\otimes {\mathcal H}_{\mr{ext}}$.  For $D \ll d$ the perturbative and non-perturbative evolutions will again agree for $t \ll \sqrt{d}$.  But  when the smallest such $D$ is instead comparable to $d$, the evolutions will generally show marked differences even for short times -- at least  in randomly-chosen initial states\footnote{In cases where there is a subspace ${\mathcal H}_{\mr{small}} \subset {\mathcal H}_{\mr{ext}}$ such that $H$ preserves the space ${\mathcal H}_{\mr{eff}} := {\mathcal H}_{\mr{p}} \otimes {\mathcal H}_{\mr{small}}$, and where the restriction of $H$ to ${\mathcal H}_{\rm{eff}}$ gives a Hamiltonian with a smaller value of $D$, the discrepancy can be reduced by choosing the initial state to lie in ${\mathcal H}_{\rm{eff}}$. }.

\subsection{Degeneracies and approximate degeneracies}
\label{sec:degen}

While the denominators in \eqref{eq:deltai2} diverge when two eigenvalues of $H$ coincide, such exact degeneracies are not in fact hard to handle.
As usual in degenerate perturbation theory, given an $n$-fold degenerate eigenspace of $H$ we may simply choose the orthonormal basis
$|\phi_1 \rangle, |\phi_2 \rangle,...|\phi_n \rangle$ in such a way that each associated $|\Delta_A \rangle$ ($A \in \lbrace 1,...,n \rbrace$) is orthogonal  to  all $X|\phi_B \rangle$ with $B \in  \lbrace 1,...,n \rbrace$ and not just for $B=A$.   This choice removes all divergent denominators from \eqref{eq:deltai2} so that the analysis of section \ref{sec:longtimegeneric} continues to apply.

The case where eigenvalue differences $\lambda_A-\lambda_B$ within each band become small as $d\rightarrow \infty$ is difficult to analyze in general.  This case also seems rather artificial as the perturbative Hamiltonian by definition knows nothing about the non-perturbative corrections that set the dimension of $\mathcal H_{\mr{np}}$ to $d$.

However, we mention briefly that we can in fact address the case where $H$ has bands of eigenvalues within each of which the
eigenvalue differences $\lambda_A-\lambda_B$ become small as $d\rightarrow \infty$ but where distinct bands remain separated by $O(1)$ gaps. In this case we can  we write $H$ as the sum of an exactly-degenerate Hamiltonian $H_0$ (whose eigenvalues label the above bands) and a term $V$ that lifts the degeneracies but leaves the splittings small.  If we assume the small splittings to be of order $1/\sqrt{d}$, then $V$ is also of order $1/\sqrt{d}$.  As a result, up to times $t \ll \sqrt{d}$, $H_0$ gives a good approximation to the evolution given by $H$.  Furthermore, the corresponding non-perturbative Hamiltonians $\tilde H_0 = XH_0X^\dagger$ and $\tilde H = XHX^\dagger$ will also generate nearly-identical time evolutions for times $t\ll \sqrt{d}$. And since we have just seen that the  analysis of section \ref{sec:longtimegeneric} can be applied to $H_0$ and $\tilde H_0$, it follows that the perturbative and non-perturbative evolutions $E_1,E_2$ defined by $H$ also agree well for times $t = O(d^\alpha)$ for $\alpha < \frac{1}{2}$.

\subsection{Finely-tuned states with large discrepancies}

An important caveat in the above analysis is that the high probability of  long-time agreement between $E_1(t)$ and $E_2(t)$ derived in section \ref{sec:longtimegeneric} holds only for what we called {\it generic} states.  These were defined to be states chosen independently of the $\tilde \alpha$-sector drawn from the Wishart distribution; i.e., the state was chosen without knowledge of the random variable $X$.  However, as we now show, there is also a  $D$-dimensional $X$-dependent subspace of $\Hn$ on which the two operators differ significantly.  This subspace is just the image of $M$ or, equivalently, the image of $X$.  Indeed, let us consider the space spanned by the $X |\phi_A \rangle$, where $|\phi_A\rangle$ again represent the eigenvectors of $H$ with non-vanishing eigenvalues. It was already argued in section \ref{sec:Preliminaries}  that $X |\phi_A \rangle$ is an approximate eigenvector of $\tilde{H}$ with eigenvalue $\tilde \lambda_A \approx \lambda_A$. Furthermore, in the limit $\frac{k}{d}\rightarrow \infty$, appendix \ref{sec:RP} shows that $X$ satisfies $X^\dagger X = X^2 = \sqrt{\frac{k}{d}}X$, which then yields
\begin{equation}
\label{eq:E1tune}
    E_2(t) X |\phi_A \rangle \approx \sqrt{\frac{k}{d}} e^{-i \lambda_A t} X |\phi_A \rangle.
\end{equation}

In contrast, let us consider the perturbative evolution $e^{-iHt}$.  Since $X$ is a uniformly-chosen random projection, $X |\phi_A \rangle$ will have only a tiny correlation with $|\phi_A\rangle$ in the limit $k \gg d$. In particular, in that limit $X |\phi_A \rangle$ will have amplitude $1$ to be orthogonal to $|\phi_A\rangle$.  And since $k\gg d \gg D$ it will also have amplitude $1$ to lie in $\ker H$. Thus,
\begin{equation}
\label{eq:E2tune}
    E_1(t) X |\phi_A \rangle \approx \sqrt{\frac{k}{d}} X |\phi_A \rangle.
\end{equation}
The two evolutions thus differ by a large phase. Of course, a global phase is of no physical relevance. However, different $X |\phi_A \rangle$ clearly lead to different phases, so physical differences between $E_1$ and $E_2$ become large for superpositions of these states.

\begin{figure}
    \centering
        \includegraphics[width=0.6\linewidth]{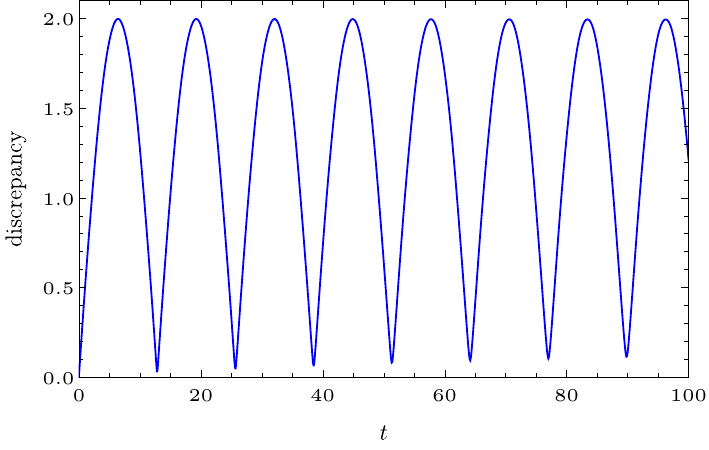}
    \caption{The figure shows the large discrepancy \eqref{eq:discr} that arises for a finely-tuned state $M|\phi_1\rangle$ with parameters $d=10^3,k=10^6,D=4$. Here the eigenvalues of the perturbative Hamiltonian $H$ are
    $\lambda_1=0.5,\ \lambda_2=1.5,\ \lambda_3=2.5,\ \lambda_4=3.5.$  The spacing between the peaks agrees well with $2\pi/\lambda_1 = 4\pi$ as predicted by comparing \eqref{eq:E1tune} with \eqref{eq:E2tune}.}
    
    \label{fig:Finely-tuned2}
\end{figure}

We thus conclude that $E_2(t) - E_1(t)$ is not small on the subspace spanned by the $X|\phi_A\rangle$. A certain caution is thus needed in distinguishing results that hold with high probability for states chosen independently of $X$ from those that hold for all possible states.  We have included a numerical example in figure \ref{fig:Finely-tuned2}.  We may also read this result backwards to say that, for any state $|\psi\rangle$, there are $\alpha$-sectors in which $(E_1(t) - E_2(t)) |\psi \rangle$ is large, namely those for which we have
\begin{equation}
    |\psi \rangle = \frac{X |\psi_1 \rangle}{\sqrt{\langle \psi_1| X^\dagger X |\psi_1 \rangle}}
\end{equation}
for some state $|\psi_1\rangle$.

\section{Discussion}
\label{sec:Discussion}
The above work investigated the effect of null states on time-evolutions defined by gravitational path integrals.  We worked in a simple topological model of quantum gravity with $k$ end-of-the-world branes, and we focused on the one-boundary sector of such models.  Since the model is topological, the source-free Hamiltonian vanishes identically.  However, general time-dependence was introduced by adding boundary sources that insert one-boundary operators defined by annihilating and creating end-of-the-world branes. Such insertions led to a non-vanishing Hamiltonian $H$ at the perturbative level, as well as a non-perturbative Hamiltonian $\tilde H$. In this context, even for times that are short in comparison with natural timescales of the perturbative Hamiltonian $H$, we found the perturbative and non-perturbative time evolutions defined by such path integrals to have large corrections when the rank $D$ of $H$ is of the same order as the dimension $d$  of the non-perturbative Hilbert space, and to be starkly different for $D\gg d$.
On the other hand, for $D \ll d$ the perturbative and non-perturbative evolutions are nearly identical for all times $t$ of order $d^\alpha$ with $\alpha < \frac{1}{2}$.  We emphasize that such timescales are exponentially large in the entropy of the system.
In particular, for $D \ll d$ we showed that the operation $E_1(t)$, defined by applying the perturbative time-evolution to a perturbative state and then mapping the result to the non-perturbative Hilbert space, was an excellent approximation to the operation $E_2(t)$ defined by first mapping the perturbative state to the non-perturbative Hilbert space and then evolving with non-perturbative time evolution.  
Our formalism also allowed us to discuss dynamics that couples our topological gravity system to a non-gravitational `bath' system, in which case analogous results apply.

A critical point is that, even though $E_1(t)$ and $E_2(t)$ differ strongly for $D \gtrsim d$,  the non-perturbative dynamics is manifestly unitary for any value of $D$.  Furthermore, this remains true when the system is coupled to an external bath.   On the other hand, section 7.3 of \cite{Guo:2021blh} argued in a related context that interactions with a system with null states was necessarily non-unitary.  The scenario considered by \cite{Guo:2021blh} involved a large parent universe that emits a set of baby universes, and where the pieces to be emitted are in a state that -- if emitted as baby universes -- would become a null state.  Thus the emission process appeared to take a state of finite-norm to one of zero norm, violating unitarity in a particularly strong way.

To resolve this tension, let us compare the discussion of \cite{Guo:2021blh} with our work above.  Since the full time evolution -- including any emissions of baby universes -- must be defined by a non-perturbative path integral, even if one described the initial state in a perturbative language we would expect the resulting operation to correspond to something like our $E_2(t)$ which continually incorporates the effects of wormholes and $\alpha$-sectors at all times $t$.  In contrast, \cite{Guo:2021blh} mentions possible effects of null states only {\it after} the baby universes have been emitted from the parent universe.  It thus assumes that the emission process itself can be described in terms of states in the perturbative Hilbert space.  It is only when the process is completed that \cite{Guo:2021blh} evaluates the result using the non-perturbative inner product.  It thus appears to us that the notion of time-evolution described in \cite{Guo:2021blh} corresponds to something like our $E_1(t)$, rather than to our $E_2(t)$. And while it is certainly true that $E_1(t)$ is generally far from unitary, we have seen that there is no tension with the unitarity of $E_2(t)$.  Indeed, our interpretation of the scenario described in \cite{Guo:2021blh} is that, since the non-perturbative evolution {\it must} remain unitary in the presence of null states, the non-perturbative time evolution is {\it required} to deviate strongly from the perturbative evolution.  Strong deviations were indeed  seen in section \ref{sec:TimeEvolLR} for our model with $D \gg d$, though of course many details remain to be understood for more realistic contexts. 

While our calculations were performed for time-independent Hamiltonians, the evolution operator $\quad$   $U(t) = \mathcal{P} \exp \left(-i \int H(t)\right)$ for time-dependent $H(t)$ can be well-approximated by a product of evolutions over short time intervals, each of which is defined by a Hamiltonian that is time-independent within the given interval.  It is thus clear that the perturbative and non-perturbative evolutions are again nearly identical over spans of time in which the non-trivial part of $H(t)$ is confined to a $D$-dimensional subspace of the perturbative Hilbert space $\mathcal H_{\mr{p}}$ with $D\ll d$, though the discrepancy should again become large once $H(t)$ explores a subspace with $D \gtrsim d$.

Although the small $D$ agreement is technically interesting, our most striking result may be the large discrepancy that arises between the perturbative and non-perturbative time evolutions for $D \gtrsim d$.  Recall, for example, that even in asymptotically AdS space
any small band of energies is associated  with an infinite set of states in the perturbative gravity description.  In particular, due to the possibility of having a black hole with a long throat deep inside the horizon, this is the case even if we require states to have no structure within a Planck distance of any event horizon (i.e., if we impose some sort of stretched horizon cutoff).  One might thus attempt to model such settings by taking the $k\rightarrow \infty$ limit of our results. Furthermore, since all such states have similar non-zero energies $E$, it is clear that the perturbative Hamiltonian has full-rank on this space.
But path integral computations (see e.g. \cite{Penington:2019kki,Almheiri:2019qdq,Marolf:2020xie,Marolf:2020rpm,Balasubramanian:2022gmo,Balasubramanian:2022lnw})  indicate that the non-perturbative theory in such contexts has only a finite number of states $d=e^{S_{\mr{BH}}}$, where $S_{\mr{BH}}$ is the associated Bekenstein-Hawking entropy.  Carrying over our results directly would then suggest that the non-perturbative dynamics in such contexts bears no relation whatsoever to the perturbative dynamics, and that it is instead essentially random.

Although our model showed no hints that quantum extremal surfaces are important for the above effect, one might nevertheless hope that in interesting gravitational theories any such effects may be confined to the interior of black holes (or, perhaps, to the regions inside and just outside an event horizon in the context described by \cite{Bousso:2023kdj}).  Furthermore, the random dynamics seen in our model appears similar in spirit to the idea that there may be no coherent semiclassical spacetime inside a black hole associated with the firewall hypothesis of \cite{Almheiri:2012rt}.  But our model suffers from an especially extreme form of such effects which appear to arise equally for `young' black holes\footnote{I.e., even for special states confined to tiny sectors of the Hilbert space which we again choose without knowledge of the random draw $M$ from the Wishart ensemble.} as well as those that are `old.'   The reader may thus naturally find such a scenario to be unpalatable.

\begin{figure}
    \centering
    \includegraphics[scale=0.7]{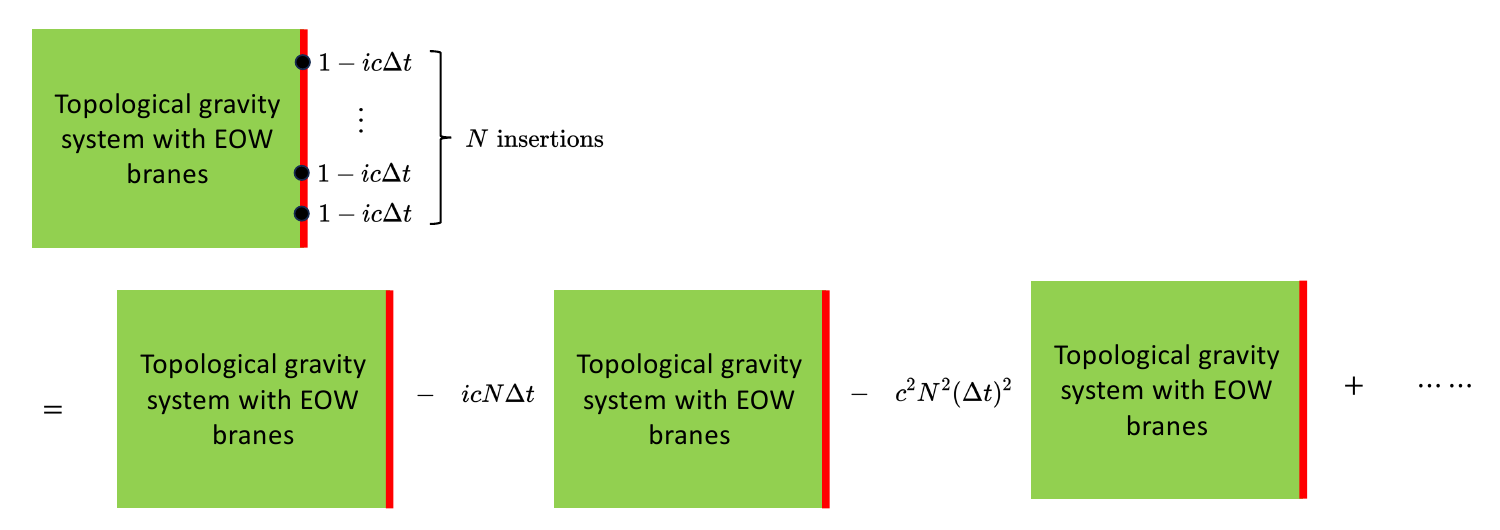}
    \caption{$N$ insertions of the c-numbers $(1-ic\Delta t)$ are equivalent to summing a series of path integrals with trivial insertions and non-trivial weights. }
    \label{fig:H-identity}
\end{figure}

A potential clue to how this predicament might be avoided can be found by recalling that we induced time-evolution in our model by inserting operators, and that we found the above results even for insertions that give $H= c {\mathds 1}_{\mr{p}}$ where $c \in {\mathds R}$ and  ${\mathds 1}_{\mr{p}}$ is the identity on the perturbative Hilbert space.  However, for this special case there is an alternative way to insert the operator $iH\Delta t$, which is to simply add a term with a single asymptotic boundary weighted by $ic\Delta t$ as shown in the 2nd term on the right in figure \ref{fig:H-identity}. For past boundary condition $|I\rangle$ and future boundary condition $|J\rangle$, the perturbative evaluation of the one-insertion path integral then yields precisely $ict\langle I| {\mathds 1}_{\mr{p}}|J\rangle_{\mr{p}} = ict\langle I|J\rangle_{\mr{p}}$ as desired.  However, instead of the matrix elements of a random operator implied by the results found in section \ref{sec:TimeEvol}, it is immediately clear that the one-insertion non-perturbative path integral computes precisely
$ict\langle \tilde I|\tilde J\rangle_{\mr{np}}$.  The corresponding non-perturbative Hamiltonian is thus $\tilde H= c {\mathds 1}_{\mr{np}}$, which is now distinctly different from the operator $XHX^\dagger$.

This, then, suggests that it is important to discuss such issues directly in the language of boundary conditions rather than using the language of operators and states on the perturbative Hilbert space.  We will return to this point in a forthcoming work.  However, for the moment we simply observe that a given model may feature certain boundary conditions for which the perturbative and non-perturbative evolutions agree to good precision for long times and which we may call np-stable.  In our topological model, the np-stable boundary conditions include the trivial boundary condition (representing the identity) and boundary conditions describing the annihilation and creation of small numbers of EOW-branes.  A tantalizing fantasy is that for higher-dimensional AdS/CFT models the np-stable boundary conditions might include those with sufficiently smooth boundary-sources of low conformal dimension (i.e., sources that define relevant or marginal deformations).  In this context, the long-time agreement we found for low-rank Hamiltonians may naturally generalize to similar successes for linear combinations of a small number of np-stable boundary conditions.  However, unless the boundary condition defines a Hamiltonian that approximates a linear combination of the identity operator and an operator of small rank, the potentially-vast difference in dimension between the perturbative and non-perturbative Hilbert spaces makes it unlikely for the perturbative and non-perturbative evolutions to act similarly across the entire space of available states.  In particular, the recurrence time for the non-perturbative evolution will generally be much shorter.  One might thus expect that a useful notion of np-stability  applies only when evolving a small and special space of states which might, for example, be associated with appropriately `young' black holes.  We hope to return to this issue in future work as well by studying non-perturbative time evolution in more realistic models, perhaps based on recent analyses  of non-perturbative geodesic lengths in JT gravity \cite{Iliesiu:2024cnh}.

As a final comment, we note that the reader may find it tempting to associate the special np-stable boundary conditions with the notion of `simple operators' described by Akers, Engelhardt, Harlow, Penington, and Vardhan in their discussion \cite{Akers:2022qdl} of non-isometric codes.  While it would certainly be interesting to explore such connections, we emphasize that our proposed notion of np-stable boundary conditions is intended to mean that the perturbative and non-perturbative path integrals both define {\it linear} operators with compatible time evolutions (perhaps on a small linear space of states).  

In contrast, the non-isometric code paradigm of \cite{Akers:2022qdl} would in principle allow us to declare in our topological model that the identity operator remains simple when represented in the form $H= \sum_I |I\rangle\ {}_{\mr{p}}\langle I|$, and to then recover the associated (trivial) perturbative dynamics from the action of the operator $XHX^\dagger$ discussed in section \ref{sec:TimeEvol}.  A similar phenomenon occurs in the dynamical model of \cite{Akers:2022qdl} where the non-perturbative evolution is simply reinterpreted as a non-isometric map applied to the desired perturbative evolution. While the non-isometric codes of \cite{Akers:2022qdl} are generally constrained by the extrapolate dictionary, such constraints become trivial in models like the one studied above that lack propagating bulk fields. And even in more complete models, this dictionary imposes no obvious constraints inside the horizon.   The freedom to choose a general non-isometric map $V$ thus allows one to essentially define $V$ to be equivariant with respect to the perturbative and non-perturbative time evolutions even if the two evolutions have no common structure deep in the bulk. The resulting flexibility makes the non-isometric code paradigm extremely powerful, though at the apparent cost of making the perturbative evolution (called the effective description in \cite{Akers:2022qdl}) essentially independent of the fundamental non-perturbative evolution.  We would prefer to instead see any effective description derived directly from a fundamental description, rather than have the two simply rendered consistent by fiat.  Again, we hope to return to such issues in future work.

\acknowledgments
The authors thank Geoff Penington for discussions of non-isometric codes.  The work of XD was supported in part by the U.S. Department of Energy, Office of Science, Office of High Energy Physics, under Award Number DE-SC0011702, and by funds from the University of California. The work of MK, XL, DM, and ZW was supported by NSF grant PHY-2107939 and by funds from the University of California. ZW was also supported by Air Force Office of Scientific Research under award number FA9550-19-1-036 and by the DOE award number DE-SC0015655. This research was supported in part by grant NSF PHY-2309135 to the Kavli Institute for Theoretical Physics (KITP).

\appendix
\section{Wishart distribution}\label{appendix:wishart}
This appendix summarizes our conventions regarding the Wishart distribution and performs a few calculations.  Other calculations described in the main text are similar.

\subsection{Basic definitions}
The Wishart distribution is a unitarily invariant distribution on the space of $k \times k$ hermitian linear matrices. 
Let us introduce a set of gaussian variables $a_{Ii}$ satisfying
\begin{align}
        \langle a_{Ii} \rangle = 0, \quad
        \langle a_{Ii} a_{Jj} \rangle = 0, \quad
        \langle a_{Ii}^\star a_{Jj}^\star \rangle = 0, \quad
        \langle a_{Ii}^\star a_{Jj} \rangle= \delta_{IJ}\delta_{ij},
\end{align}
where $I,J \in \lbrace 1,...,k \rbrace$ and $i,j \in \lbrace 1,...,d \rbrace$.  All higher moments of the $a_{Ii}$ may be calculated using Wick's theorem.

From the $a_{Ii}$, we may build a rectangular random matrix that maps from $\mathbb{C}^d$ to $\mathbb{C}^k$:
\begin{equation}
    [G]_{Ii} = a_{Ii}.
\end{equation}
We may then construct a random $k \times k$ matrix $M$ which is said to be drawn from the Wishart distribution:
\begin{equation}\label{eq:defM}
    M := \frac{1}{d} GG^\dagger.
\end{equation}
More explicitly, this matrix may be written
\begin{equation}
    [M]_{IJ} = \frac{1}{d} \sum_{i=1}^d a_{Ii}a^\star_{Ji}.
\end{equation}

The matrix $M$ is hermitian by construction. Moreover, because it can be expressed as the  product \eqref{eq:defM}, we have
\begin{equation}
    \text{rank}\, M \le \min (d,k).
\end{equation}
The inequality is saturated with probability one as obtaining a smaller rank requires an accidental degeneracy.

\subsection{A few moments}
We will now explicitly calculate the first few moments of $M$. This is a nice a warm-up before performing some of the more intricate calculations described in the main text.
\subsubsection{Mean}
We want to calculate $\langle M \rangle$. This is quite immediate, since we have:
\begin{equation}
%\label{eq:avM2}
    [\langle M \rangle]_{IJ} = \frac{1}{d} \sum_{i=1}^d\langle a_{Ii}a^\star_{Ji} \rangle = \frac{1}{d}  \sum_{i=1}^d\delta_{IJ} \delta_{ii} = [{\mathds 1}_{\mr{p}}]_{IJ}, \label{appA:one_point}
\end{equation}
and thus $\langle M \rangle =  {\mathds 1}_{\mr{p}}$.
\subsubsection{Variance}
To calculate the variance of $M$ we will also need $\langle M^2 \rangle$. More explicitly, we wish to compute the enesmble average of the matrix
\begin{equation}
    [M^2]_{IK} = \sum_{J=1}^k M_{IJ}M_{JK} = \frac{1}{d^2} \sum_{i,j=1}^d \sum_{J=1}^k\left(
    a_{Ii}a^\star_{Ji} a_{Jj} a^\star_{Kj}
    \right).
\end{equation}
Taking $\langle \cdot \rangle$ on both sides and applying Wick's theorem yields
\begin{equation}
    \langle M^2 \rangle_{IK} = \frac{1}{d^2}\sum_{i,j=1}^d \sum_{J=1}^k \langle a_{Ii}a^\star_{Ji} a_{Jj} a^\star_{Kj} \rangle = \frac{1}{d^2} \sum_{i,j=1}^d \sum_{J=1}^k\left(
    \langle a_{Ii}a^\star_{Ji} \rangle \langle a_{Jj} a^\star_{Kj} \rangle + \langle a_{Ii}  a^\star_{Kj} \rangle \langle a_{Jj} a^\star_{Ji} \rangle
    \right),
\end{equation}
and thus
\begin{equation}
    \langle M^2 \rangle_{IK} = \frac{1}{d^2} \sum_{i,j=1}^d \sum_{J=1}^k \left(
    \delta_{IJ} \delta_{ii} \delta_{JK} \delta_{jj} + \delta_{IK}\delta_{ij} \delta_{JJ}
    \right) =  \delta_{IK} + \frac{k}{d}\delta_{IK} = \left(
    1 + \frac{k}{d}
    \right) [{{\mathds 1}_{\mr{p}}}]_{IK}.
\end{equation}
The variance is then  $\langle M^2 \rangle - \langle M \rangle^2 = \frac{k}{d} \mathrm{{\mathds 1}_{\mr{p}}}.$ We see that for  $k \gg d$, the matrix $M$ is far from being sharply peaked.  This should be no surprise since, although the ensemble is invariant under $k\times k$ unitaries, any given $M$ has rank $d \ll k$ and must thus determine a preferred $d$-dimensional subspace of a rank $k$ vector space. 

Another useful result is the two-point function
$    \langle M_{I_1 I_2} M_{I_3 I_4} \rangle$.
Wick's contractions analogous to those above yield
\begin{equation}
    \langle M_{I_1 I_2} M_{I_3 I_4} \rangle = [{\mathds 1}_{\mr{p}}]_{I_{1} I_2} [{\mathds 1}_{\mr{p}}]_{I_3 I_4} + \frac{1}{d} [{\mathds 1}_{\mr{p}}]_{I_1 I_4} [{\mathds 1}_{\mr{p}}]_{I_2 I_3}.
    \label{appA:two_point}
\end{equation}
From \eqref{appA:two_point} we can also see that for any (fixed) $k \times k$ matrix $H$ we have:
\begin{equation}
\label{eq:MHM}
    \langle MHM \rangle = H + \frac{\text{Tr}_{\mr{p}}{H}}{d} {\mathds 1}_{\mr{p}}.
\end{equation}
In particular, taking $H = {\mathds 1}_{\mr{p}}$ recovers our previous result.

\subsection{Random projections}
\label{sec:RP}
For large $k$, the central limit theorem tells us that the magnitudes of the vectors $\sum_{I} a_{Ii} |I\rangle$ should have small fluctuations.  Furthermore, for $k\gg d$ we expect these $d$ vectors to be approximately orthonormal.  In this limit we thus expect $M$ to be proportional to a rank $d$ projection.  If this is so, then the result \eqref{eq:average} tells us that the projection must be ${\cal P} = \frac{d}{k} M$.
Since the distribution for $M$ is clearly  invariant under unitary transformations, the range of ${\cal P}$ must be a random subspace with respect to the uniform distribution\footnote{Geometrically, the space of all $d$-dimensional subspaces is $\mathrm{U}(k)/\left( \mathrm{U}(d) \times \mathrm{U}(k-d)
\right)$.  The uniform distribution on this space is given by the associated Haar measure.}.

We now explore the details of this argument.   Since ${\cal P}$ is manifestly Hermitian, it is a projection if ${\cal P}^2={\cal P}$. A straightforward yet tedious calculations yields
\begin{subequations}
\begin{equation}
    \langle {\cal P}^2-{\cal P} \rangle=\frac{d^2}{k^2}[{\mathds 1}_{\mr{p}}],
\end{equation}
while
\begin{equation}
\label{eq:projkggd}
\begin{aligned}
     \langle (\mathcal P^2 - \mathcal P)^2 \rangle&= \langle \mathcal P^4 \rangle + \langle \mathcal P^2 \rangle-2 \langle \mathcal P^3 \rangle\\
     &=\left( \frac{4d^2}{k^4}+\frac{d^4}{k^4}+\frac{3d}{k^3}+\frac{4d^3}{k^3}+\frac{2d^2}{k^2}\right)[{\mathds 1}_{\mr{p}}].
     \end{aligned}
\end{equation}
\end{subequations}

In the limit $k/d\rightarrow \infty$ we thus find that $\mathcal P$ approaches a projection in the sense of the weak operator topology. However, a stronger measure of convergence is given by the the Frobenius-norm, or the Schatten 1-norm:
\begin{equation}
    ||Y||^2 := \tr (YY^\dagger),
\end{equation}
which is easily computed by
taking the trace of \eqref{eq:projkggd}. We find
\begin{equation}
    ||\mathcal P^2 - \mathcal P||^2 \sim \frac{2d^2}{k}.
\end{equation}
So even in this stronger sense we find $\mathcal P$ to approach a projection in the stronger limit $k/d^2\rightarrow \infty$.

\cleardoublepage
\addcontentsline{toc}{section}{References}
\bibliographystyle{JHEP}
\bibliography{ref.bib}

\end{document}